\documentclass [prb,twocolumn,superscriptaddress,floatfix] {revtex4-1}
\usepackage{amsmath}
\usepackage{graphicx}
\usepackage{lmodern}
\usepackage{amsmath}
\usepackage{hyperref}
\usepackage{color}
\usepackage{amssymb}
\usepackage{bm}
\usepackage{ulem}
\newcommand{\beq} {\begin{equation}}
\newcommand{\eeq} {\end{equation}}
\newcommand{\bea} {\begin{eqnarray}}
\newcommand{\eea} {\end{eqnarray}}
\newcommand{\be} {\begin{equation}}
\newcommand{\ee} {\end{equation}}

\newcommand{\crr} {{\text{cross}}}

\DeclareMathOperator{\sgn}{sgn}

\DeclareMathOperator{\Ree}{Re}

\begin{document}
\title {The interplay between superconductivity and non-Fermi liquid  at a quantum-critical point in a metal.}
\author{Andrey V. Chubukov}
\affiliation{School of Physics and Astronomy and William I. Fine Theoretical Physics Institute,
University of Minnesota, Minneapolis, MN 55455, USA}
\author{Artem Abanov}
\affiliation{Department of Physics, Texas A\&M University, College Station,  USA}
\author{Yuxuan Wang}
\affiliation{Department of Physics, University of Florida, Gainesville, FL 32611,  USA}
\author{Yi-Ming Wu}
\affiliation{School of Physics and Astronomy and William I. Fine Theoretical Physics Institute, University of Minnesota, Minneapolis, MN 55455, USA}
\date{\today}
\begin{abstract}
  Near a quantum-critical point, a metal reveals two competing tendencies: destruction of fermionic coherence and attraction in one or more pairing channels.
We analyze the competition within Eliashberg theory for a class of quantum-critical models with an effective dynamical electron-electron interaction
 $V(\Omega_m) \propto 1/|\Omega_m|^\gamma$ (the $\gamma$-model) for $0 < \gamma <1$.
  We argue that the two tendencies are comparable in strength, yet
  the one towards pairing is stronger, and the ground state is a superconductor. We show, however, that there exist two distinct regimes of system behavior below the onset temperature of the pairing $T_p$. In the range $T_{\crr} < T < T_p$ fermions remain incoherent and the spectral function $A(k, \omega)$ and the density of states $N(\omega)$  both display ``gap filling" behavior in which,  e.g.,
  the position of the maximum in $N(\omega)$ is set by temperature rather than the pairing gap. At lower $T < T_{\crr}$, fermions acquire coherence, and $A(k, \omega)$ and  $N(\omega)$  display conventional "gap closing" behavior, when the peak position in $N(\omega)$ scales with the gap and shifts to a smaller value as $T$ increases.  We argue that the existence of the two regimes comes about because of special behavior
   of fermions with frequencies $\omega = \pm \pi T$ along the Matsubara axis. Specifically, for these fermions, the component of the self-energy, which competes with the pairing, vanishes in the normal state.  We further argue that the crossover  at $T \sim T_{\crr}$ comes about because Eliashberg equations allow an infinite number of topologically distinct solutions for the onset temperature of the pairing within the same gap symmetry. Only one solution, with the highest $T_p$, actually emerges, but other solutions are generated and  modify the form of the gap function
   at $T \leq  T_{\crr}$. Finally, we argue that the actual $T_c$ is comparable to $T_{\crr}$, while at $T_{\crr} < T < T_{p}$ phase fluctuations destroy superconducting long-range order, and the system displays   a pseudogap behavior.
\end{abstract}
\maketitle

\section{ Preface}

It is our great pleasure to present this mini-review for the special issue of Annals of Physics devoted to 90th birthday of Gerasim Matveevich Eliashberg.  His works, particularly on electron-phonon superconductivity outside the weak coupling limit, are of the highest scientific quality.  The Eliashberg theory of superconductivity is simultaneously a rigorous extension of BCS theory, controlled by a small parameter, and a tool to compute superconducting $T_c$ and observables, such as thermodynamic variables like specific heat and magnetic susceptibility, and dynamic characteristics, like the spectral function and the density of states.  In ``high -$T_c$ era'' Eliashberg theory has been extended to the cases when the pairing is of electronic origin, mediated by collective excitations in spin or charge channel. Eliashberg theory of spin-fluctuation superconductivity is a ``canonical'' topic in the studies of Cu, and Fe-based superconductors, heavy fermion superconductors, organic superconductors, and other classes of systems.  In this mini-review we summarize the efforts by several groups, including ours, to extend Eliashberg theory to the new regime when the pairing boson becomes massless. This happens, most naturally, when the system approaches an instability towards a spin or charge order.  Amazingly, Eliashberg equations in this critical regime reveal qualitatively new physics, not seen in the cases when a pairing boson has a finite mass. Still, the  works by Gerasim Matveevich were the ones which established  the solid base for all today's studies of quantum-critical metals.

We hope that this work will show our profound admiration of Gerasim Matveevich Eliashberg. We wish him the very best.

\section{ Introduction.}
Pairing near a quantum-critical point (QCP) in a metal is a fascinating subject, which attracted quite substantial attention in the correlated electron community after the discovery of superconductivity in heavy fermion and organic materials, in the cuprates, and, more recently, in Fe-pnictides and Fe-chalcogenides.
\cite{combescot,
Bergmann,
Bergmann2,
ad,
Marsiglio_88,
Marsiglio_91,
Karakozov_91,
nick_b,
acf,
acs,
acs2,
finger_2001,
acn,
pepin,
q01,
q02,
2kf,
2kf2,
2kf3,
son,
son2,
sslee,
sslee2,
subir,
subir2,
moon_2,
max,
max2,
senthil,
max_last,
raghu,
raghu2,
raghu3,
raghu4,
raghu5,
scal,
scal2,
book1,
review,
review2,
review3,
review4,
efetov,
wang,
wang2,
raghu_15,
Wang2016,
steve_sam,
tsvelik,
tsvelik_1,
tsvelik_2,
tsvelik_3,
khodas,
vojta,
mack,
varma,
Rice,
matsuda,
metzner,
metzner_1,
metzner_2,
berg,
berg_2,
berg_3,
kotliar,
kotliar2,
review3,
tremblay_2,
georges,
georges2,
khvesh,
Kotliar2018,
we_last_D,
wu_1, abanov_last,
avi_1,
avi_2, schmalian_19,wang_19,schmalian_19a}
QC itinerant models, analyzed in recent years,  include fermions in spatial dimensions $D \leq 3$ at the verge of either spin-density-wave (SDW) or charge-density-wave instability, near an instability towards $q=0$ Pomeranchuk order in spin-or charge channel (a nematic QCP),
 2D fermions on a half-filled Landau level, and color superconductivity of quarks, mediated by gluon exchange.
  Very recently, the list has been extended to several SYK-type models with either electron-electron or electron-phonon interaction (see the article by Daniel Hauck, Markus Klug, Ilya Esterlis, and J\"{o}rg Schmalian for this issue).

From the theoretical perspective, the  key interest in the pairing near a QCP is due to the fact that an effective electron-electron interaction, mediated by a critical collective boson, which condenses on one side of a QCP, provides strong attraction in one or more pairing channels and therefore acts as a stronger glue for superconductivity (SC) than electron-phonon interaction.  The same effective interaction,  however, also  gives a singular contribution to the fermionic self-energy and thus tends to make fermions incoherent and gives rise to non-Fermi liquid (NFL) physics.
  The two tendencies compete with each other: fermionic incoherence destroys Cooper logarithm and reduces the tendency to pairing,  while the opening of a superconducting gap eliminates the scattering at  low energies  and reduces the tendency to NFL behavior.
  To find the outcome of the interplay between SC and NFL,
one needs to analyze  the set of coupled integral equations for the fermionic self-energy on the FS $\Sigma ({\bf k}, \omega)$ and the pairing vertex $\Phi ({\bf k}, \omega)$ for fermions with $({\bf k}, \omega)$ and $(-{\bf k}, - \omega)$.  Equivalently, one can analyze the equations for the inverse quasiparticle residue $Z ({\bf k}, \omega) = 1 + \Sigma({\bf k}, \omega)/\omega$ and the gap function $\Delta ({\bf k}, \omega) = \Phi ({\bf k}, \omega)/Z({\bf k}, \omega)$.

  We consider the subset of models in which collective bosons are slow modes compared to dressed fermions. In this situation, one can analyze the interplay between NFL and SC by extending the Eliashberg theory for
    electron-phonon interaction  to the case of pairing due to electron-electron interaction.
     Within Eliashberg theory, the self-energy and the pairing vertex can be approximated by their values at the Fermi surface (FS).    The self-energy on the FS, $\Sigma ({\bf k}, \omega)$ is invariant under rotations from the point group of the underlying lattice. The angular variation of the gap function  $\Delta ({\bf k}_F, \omega)$ and  relative phases of  $\Delta ({\bf k}_F, \omega)$ on different FS's in multi-band systems
      are model specific. Near a ferromagnetic QCP, the strongest attraction is in the $p$-wave channel.
      Near an antiferromagnetic  QCP, the strongest  is in $d-$wave channel in the case when there is a single FS, and the largest density of states (DOS) is around $(0,\pi)$  and symmetry related points, as in the cuprates. In the same geometry,
      near a QCP towards a CDW order with a small $q$, superconductivity can be either $s-$wave or $d-$wave.  For nearly compensated metal with hole and electron pockets, as in Fe-based superconductors, the two attractive channels near a SDW QCP are $s^{+-}$ and $d-$wave.  Near a $q=0$ nematic QCP, the pairing vertex is peaked at the FS points, where the form-factor in the corresponding particle-hole channel is at maximum,
and superconductivity mediated by nematic fluctuations can be
  $s-$wave, $p-$wave, $d-$wave, etc.
    In each  case one has to project the pairing interaction into the proper irreducible channel and solve for the pairing vertex with a given symmetry.
   In principle, even after projection one has to solve integral equation in  momentum space as in a lattice system
    each irreducible representation contains an infinite set of eigenfunctions.
    However, in the two limiting cases when either one of these eigenfunctions gives the dominant contribution to the gap (e.g., $\cos k_x - \cos k_y$ for $d-$wave pairing in the cuprates, compared to $\cos{(2m+1) k_x} - \cos{(2m+1) k_y}$
  with all other $m$'s), or all eigenfunctions are relevant
  (but the pairing is confined to a narrow range on the FS around ``hot spots"),  the momentum integration can be carried out exactly for the pairing vertex and the self-energy.  In this situation,  the original set of coupled equations for the self-energy and the pairing vertex in $D$ spatial dimension and one time dimension reduces to the set of coupled 1D equations for frequency-dependent
    $\Sigma (\omega)$ and $\Phi (\omega)$, with frequency-dependent interaction $V(\Omega)$.

    Away from a QCP,  $V(\Omega)$ tends to a finite value at $\Omega =0$. Then fermionic self-energy has a FL form at the smallest frequencies, and the pairing kernel is
     logarithmically singular, as in BCS theory.   Then already an infinitesimally small attraction gives rise to superconductivity.   At larger $\Omega$, the pairing interaction decreases, which implies that the frequency integrals for the self-energy and the pairing vertex are infra-red convergent. The same behavior at small and large $\Omega$ holds for $V(\Omega)$ due to phonon scattering, and the analysis of electronically-mediated superconductivity
      away from a QCP  is almost identical to Eliashberg theory for phonon-mediated superconductivity, the only  distinction is that for electronically-mediated pairing, $V(\Omega)$ by itself changes below $T_c$.
      At a QCP, the situation is qualitatively different as the interaction $V(\Omega)$, mediated by a critical boson, diverges at $\Omega =0$ as  $V(\Omega_m) \propto 1/\Omega^\gamma$.  The exponent $\gamma>0$ depends on the model, ranging from small $\gamma =
 O(\epsilon)$ in models in $D = 3 -\epsilon$ to $\gamma \leq 1$ in 2D models at SDW, CDW, and nematic QCP.   Besides these examples of electronically-mediated pairing, the case $\gamma =2$ corresponds to fermions interacting with an Einstein phonon, in the (properly defined) limit of vanishing Debye frequency. The model with $V(\Omega) \propto 1/\Omega^\gamma$ has been nicknamed the $\gamma-$ model, and we will use this notation.

\subsection{Brief summary of the results and the structure of the paper}

In this paper we consider the system behavior for $0 <\gamma <1$.  The analysis for larger $\gamma >1$ is more involved and requires separate consideration. We show that a  non-FL self-energy in the normal state does not prevent the formation of bound states of fermions with opposite momenta and frequency.  We argue, however, that there exist two distinct regimes of system behavior below the onset temperature of the pairing $T_p$.
 Immediately below $T_p$, down to some finite temperature $T_{\crr}$,
  the pairing does not change qualitatively the fermionic self-energy, which retains its non-Fermi liquid form. As the consequence, fermions remain incoherent. We show that in this $T$ range the DOS $N(\omega)$  displays $\omega/T$ scaling and "gap filling" behavior, meaning that the position of the maximum in $N(\omega)$ is set by temperature rather than the pairing gap. The spectral function $A(k, \omega)$ displays either the same "gap filling" behavior as $N(\omega)$ or "Fermi arc" behavior, depending on the type of the pairing and the position of ${\bf k}$ along the FS.   At lower $T < T_{\crr}$ fermions acquire coherence due to feedback from gap opening, and
  $N(\omega)$  displays a BCS-like "gap closing" behavior, in which the peak position in $N(\omega)$ scales with the gap $\Delta (T)$ and shifts to smaller value as $T$ increases and the gap gets smaller.  The spectral function also behaves as expected for a BCS superconductor. The crossover temperature $T_{\crr}$  roughly corresponds to $\Delta (T_{\crr}) = T_{\crr}$.

    We argue that the existence of the two regimes comes about because of special behavior
   of fermions with Matsubara frequencies $\omega = \pm \pi T$. Specifically, for these fermions, the component of the self-energy, which competes with the pairing, vanishes in the normal state~\cite{Wang2016}. As the consequence, strong pairing interaction between fermions with $\omega = \pi T$ and $\omega = -\pi T$ is not counter-weighted by NFL self-energy.  We show that, immediately below $T_p$, the pairing gap for fermions with all other Matsubara frequencies does not develop on its own, but rather is induced by the opening of the gap for fermions with $\omega_m = \pm \pi T$. In this situation,  $\Delta (\omega_m)$  is strongly peaked at $\omega_m = \pm \pi T$.  This gives rise to $\omega/T$ scaling in real frequencies and to "gap filling" behavior (Ref. \cite{wu_1}

     We  argue that the crossover to BCS-like behavior  at $T \sim T_{\crr}$ comes about because Eliashberg equations at a QCP  allow an infinite number of topologically distinct solutions for the onset temperature of the pairing within the same gap symmetry~\cite{WAWC}. Only one solution, with the highest $T_p$, actually emerges (the one induced at $T_p$ by fermions with $\omega_m =\pm \pi T$). However, below $T_p$, when the actual $\Delta (\omega_m)$ is the
       solution of the non-linear Elishberg equation,  other gap components get generated due to non-linear coupling between different solutions within the same pairing symmetry.
        This gives rise to a  modification of the form of $\Delta (\omega_m)$, which becomes less peaked at $\omega_m = \pm \pi T$. The  modification becomes strong
      at around $T_{\crr}$, and at smaller $T$ fermions with all Matsubara frequencies equally contribute to pairing. This, we argue, gives rise to BCS-like behavior.

        Finally, we argue that  in the range $T_{\crr} < T < T_p$  superfluid stiffness $\rho_s$ is smaller than  $T$ (Ref. \cite{abanov_last}. In this situation, phase fluctuations likely destroy superconducting long-range order. At smaller $T< T_{\crr}$, the stiffness is much larger, of order $\Delta$ (it would be of order $E_F$ if a pairing boson was massive, with sufficiently large mass).  In this situation, it is natural to expect that the actual $T_c$ is comparable to $T_{\crr}$, while in between $T_{\crr}$ and $T_p$ the system displays a pseudogap behavior.

  The paper is organized as follows. In Sec.~\ref{sec:model} we briefly review the $\gamma$ model with effective fermion-fermion interaction mediated by a gapless boson with $V (\Omega_m) = (g /|\Omega_m|)^\gamma$ and present Eliashberg equations for our case.
    In Sec.\ref{Tp} we show the results of numerical solution of the linearized equation for the pairing vertex (or the gap function), which determines the onset temperature for the pairing $T_p = T_p (\gamma)$.
   In Sec. \ref{largeN} we
   extend the $\gamma$ model to make
     the interaction in the particle-particle channel relatively  smaller by the factor $1/N$, where $N>1$.
         In Sec. \ref{finiteT_1} we discuss the
     solution
     of the full non-linear Eliashberg equations at a finite $T$ below $T_p$, identify two different types of system  behavior at larger and smaller $N$, and show that the crossover temperature between the two regimes, $T_{\crr} (N)$ terminates at some $N_{cr} >1$
      In Sec. \ref{T=0} we present the results of the analytical study of Eliashberg equations at $T=0$, which show that $N_{cr}$ indeed exists and
       separates the NFL ground state
        at $N > N_{cr}$ and the  SC state
         at $N < N_{cr}$. Here we argue that at $N < N_{cr}$
         there is an infinite discrete set of solutions for the pairing gap, $\Delta_n (\omega)$, ranging from the BCS-type solution to the solution with infinitesimally small gap.
      In Sec. \ref{finite_T_2} we show that each solution from the set at $T=0$ evolves with $T$ and ends up at its own critical temperature $T_{p,n}$.  The BCS-like solution ($n=0$) ends at $T_{p,0} = T_p$, which we found before. Other solutions end at smaller $T_{p,n}$.
      In Sec. \ref{final} we combine our results and present our understanding of the crossover at $T = T_{\crr}$.
          In Sec.~\ref{rho} we  present the results for the superfluid stiffness $\rho_s (T)$ and argue
            that
          the actual $T_c \sim T_{\crr}$, while at $T_{\crr} < T < T_p$ the system displays pseudogap behavior.
          In Sec. \ref{sec:application} we briefly compare our results for the spectral function with ARPES data for the cuprates.
We present the summary of our
results in Sec. \ref{sec:summary}.

\section{The model}
\label{sec:model}

We  consider a model of itinerant fermions  at the onset
of a  long-range order in either spin or charge channel.  At the critical point the propagator of a soft boson
 becomes massless and
 mediates singular interaction between fermions. We follow earlier works~\cite{acf,acs,moon_2,max,senthil,scal,efetov,max_last,raghu_15,haslinger,Wang2016,Kotliar2018,wu_1,abanov_last}  and assume that this interaction is attractive in at least one pairing channel and that a pairing
  boson can be treated as slow mode compared to a fermion, i.e., at a given momentum $q$, typical fermionic frequency is much larger than  typical bosonic frequency. This is the case  for a conventional phonon-mediated superconductivity, where
    for $q \sim k_F$ a typical fermionic frequency is of order $E_F$, while typical bosonic frequency is of order Debye frequency $\omega_D$.  The ratio  $\delta_E =\omega_D/E_F$ is the  small parameter for Eliashberg
     theory of phonon-mediated superconductivity. This theory allows one to obtain a set of coupled integral equations for frequency dependent fermionic self-energy and the pairing vertex.  By analogy,  the theory of electronic superconductivity, mediated by  soft collective bosonic excitations in spin or charge channel, is also often called Eliashberg theory.  We will use this notation.

   Justification of Eliashberg theory for electronically mediated superconductivity is case specific and sometimes a small parameter for Eliashberg  approximation can be found only by extending a model e.g., to a large number of fermionic flavors.  Furthermore, for several 2D models, the corrections to Eliashberg approximation for the self-energy  in the normal state are logarithmically singular and in the absence of the pairing would change the system behavior at the smallest frequencies.   Here we assume that the onset temperature for the pairing, $T_p$, is larger, at least numerically, than the scale at which corrections to Eliashberg approximation become relevant, and stick with the Eliashberg theory.

Within the Eliashberg  approximation,
 one can explicitly integrate over the momentum component perpendicular to the Fermi surface (for a given pairing symmetry) and reduce the
   pairing problem  to a set of coupled integral equations for frequency dependent self-energy $\Sigma (\omega_m)$
   and the pairing vertex $\Phi (\omega_m)$ with effective frequency-dependent dimensionless interaction $\chi(\Omega) = (g/|\Omega|)^\gamma$.   This interaction gives rise to NFL form of the self-energy in the normal state and, simultaneously,
    gives rise to the pairing.

  The Eliashberg equations are
    \bea
    &&\Phi (\omega_m) =
    \pi T  g^\gamma \sum_{m'} \frac{\Phi (\omega_{m'})}{\sqrt{{\tilde \Sigma}^2 (\omega_{m'}) +\Phi^2 (\omega_{m'})}}
    ~\frac{1}{|\omega_m - \omega_{m'}|^\gamma}, \nonumber \\
     &&{\tilde \Sigma} (\omega_m) = \omega_m \nonumber\\
   &&+  g^\gamma
    \pi T \sum_{m'}  \frac{{\tilde \Sigma}(\omega_{m'})}{\sqrt{{\tilde \Sigma}^2 (\omega_{m'})  +\Phi^2 (\omega_{m'})}}
    ~\frac{1}{|\omega_m - \omega_{m'}|^\gamma} \label{eq:gapeq}
\eea
 where here and below ${\tilde \Sigma}(\omega_{m}) = \omega_m + \Sigma (\omega_m)$.
  Note that we define $\Sigma (\omega_m)$ as a real function of frequency, i.e., without the overall factor of $i$.

  The  superconducting gap $\Delta (\omega_m)$ is defined as a real variable
\beq
 \Delta (\omega_m) = \omega_m  \frac{\Phi (\omega_m)}{{\tilde \Sigma} (\omega_m)}
  \label{ss_1}
  \eeq
   The equation for $\Delta (\omega)$ is readily obtained from (\ref{eq:gapeq}):
   \beq
   \Delta (\omega_m) = \pi T  g^\gamma \sum_{m'} \frac{\Delta (\omega_{m'}) - \Delta (\omega_m) \frac{\omega_{m'}}{\omega_m}}{\sqrt{\omega^2_{m'} +\Delta^2 (\omega_{m'})}}
    ~\frac{1}{|\omega_m - \omega_{m'}|^\gamma}.
     \label{ss_11}
  \eeq
   This equation contains a single function $\Delta (\omega)$, but  for the price that $\Delta (\omega_m)$ appears on both sides of the equation, which makes (\ref{ss_11}) less convenient for the analysis than Eqs. (\ref{eq:gapeq}).

The full set of Eliashberg equations for electron-mediated pairing contains also the equation describing the feedback from the pairing on $\chi(\Omega)$, e.g., the emergence of a propagating mode (often called a resonance mode)  in the dynamical spin susceptibility for $d-$wave pairing mediated by antiferromagnetic spin fluctuations~~\cite{abanov_ch,eschrig}.  To avoid additional complications, we do not include this feedback into our consideration.  In general terms, the feedback from the pairing makes bosons less incoherent and can be modeled by assuming that the exponent
  $\gamma$ moves towards larger value as $T$ moves down from $T_p$.

The two equations in  (\ref{eq:gapeq}) describe the interplay between two competing tendencies -- the tendency towards superconductivity, specified by $\Phi$,  and the tendency towards incoherent NFL behavior, specified by $\Sigma$.  The competition between the two tendencies is encoded in the fact that $\Sigma$ appears in the denominator of the equation for $\Phi$ and $\Phi$ appears in the denominator of the equation for $\Sigma$.
 Accordingly, a large, non-FL self-energy is an obstacle to Cooper pairing, while once $\Phi$  develops, it reduces the strength of the self-energy, i.e., moves a system back into a FL regime.

 As we said in the Introduction, Eqs. (\ref{eq:gapeq})-(\ref{ss_11})  describe color superconductivity~\cite{son,son2} and pairing in 3D ($\gamma = 0_+$, $\chi (\Omega_m) \propto \log{|\omega_m|}$),  spin- and charge-mediated pairing in $D=3-\epsilon$ dimension~\cite{senthil,max_last,raghu_15} and superconductivity in graphene~\cite{khveshchenko} ($\gamma = O(\epsilon) \ll 1$),  a 2D pairing ~\cite{2kf}  with  interaction peaked at $2k_F$ ($\gamma =1/4$),   pairing at a 2D nematic/Ising-ferromagnetic QCP~\cite{nick_b,steve_sam,triplet,*triplet2,*triplet3} ($\gamma =1/3$),   pairing at a 2D $(\pi,\pi)$ SDW QCP~\cite{acf,acs,millis_05,wang} and an incommensurate CDW QCP~\cite{ital,*ital2,*ital3,wang_2,*wang_22,*wang23} ($\gamma =1/2$),
   dispersionless fermions randomly interacting with an Einstein 
   ~\cite{wang_19,schmalian_19,schmalian_19a}  and a spin-liquid model for the cuprates~\cite{tsvelik} ($\gamma =0.7$)
  a 2D pairing  mediated by an undamped  propagating boson ($\gamma =1$),  pairing in several Fe-based superconductors~\cite{kotliar} ($\gamma =1.2$) and even
  the strong coupling limit of phonon-mediated superconductivity for either dispersion-full~\cite{combescot,Bergmann,*Bergmann2,*ad,Marsiglio_88,*Marsiglio_91,Karakozov_91} or dispersion-less~\cite{schmalian_19} fermions ($\gamma =2$).
  The pairing models with parameter-dependent $\gamma$ have been analyzed as well (Refs. \onlinecite{subir,moon_2}).   The case $\gamma =0$ describes a BCS superconductor.
  Here we consider the set of $\gamma$-models with  $\gamma < 1$.

   The r.h.s. of the equations for $\Phi (\omega_m)$ and $\Sigma (\omega_m)$ contain  divergent
   contributions
    from the terms with $m' =m$, i.e., from $\chi (0)$.
    The divergence can be regularized by moving slightly away from a QCP, in which case
   $\chi (0)$ is large but finite.  This term mimics the effect of non-magnetic impurities and by Anderson theorem should not affect $T_p$.  To get rid of
   this thermal contribution
   in the equations for $\Phi (\omega)$ and $\Sigma (\omega)$, we follow Refs.~\cite{msv,acn} and use the same trick as in the derivation of the Anderson theorem~\cite{agd}.
    Namely, in each equation in (\ref{eq:gapeq}) we  pull out the term with $m'=m$ from the summand and move it to the l.h.s.. We then introduce new variables $\Phi^* (\omega_m)$ and $\Sigma^* (\omega_m)$ as
   \bea
   \Phi^* (\omega_m) &=& \Phi (\omega_m)
   \left(1- Q (\omega_m)\right), \nonumber \\
   {\tilde \Sigma}^* (\omega_m) &=& {\tilde \Sigma} (\omega_m) \left(1- Q (\omega_m)\right)
   \label{ss_2}
   \eea
    where
    \beq
     Q (\omega_m) =   \frac{\pi T \chi (0)}{\sqrt{{\tilde \Sigma}^2 (\omega_{m}) +\Phi^2 (\omega_{m})}}
   \label{ss_2_a}
   \eeq
    The ratio $\Phi (\omega_m)/ {\tilde \Sigma} (\omega_m) = \Phi^* (\omega_m)/ {\tilde \Sigma}^* (\omega_m)$, hence $\Delta (\omega_m)$, defined in (\ref{ss_1}),  is invariant under $\Phi (\omega_m) \to \Phi^* (\omega_m)$ and ${\tilde \Sigma} (\omega_m) \to {\tilde \Sigma}^* (\omega_m)$.  Using  (\ref{ss_2}), one can easily verify that the equations on $\Phi^* (\omega_m)$ and ${\tilde \Sigma}^* (\omega_m)$  are the same as in (\ref{eq:gapeq}), but without the thermal
    contribution, i.e., the summation over $m'$ now excludes the divergent term with $m' =m$.
        In the gap equation, the term with $m = m'$
      vanishes  because the vanishing of the  numerator in the r.h.s. of (\ref{ss_11}).

The  equations for $\Phi^* (\omega_m)$ and ${\tilde \Sigma}^* (\omega_m)$ are
 \bea
   && \Phi^* (\omega_m) =\nonumber\\
&&
    \pi T  g^\gamma \sum_{m' \neq n} \frac{\Phi^* (\omega_{m'})}{\sqrt{({\tilde \Sigma}^* (\omega_{m'}))^2 +(\Phi^* (\omega_{m'}))^2}}
    ~\frac{1}{|\omega_m - \omega_{m'}|^\gamma}, \nonumber \\
   &&  {\tilde \Sigma}^* (\omega_m) = \omega_m +\nonumber\\
   &&  g^\gamma
    \pi T \sum_{m' \neq m}  \frac{{\tilde \Sigma}^* (\omega_{m'})}{\sqrt{({\tilde \Sigma}^* (\omega_{m'}))^2  +(\Phi^* (\omega_{m'}))^2}}
    ~\frac{1}{|\omega_m - \omega_{m'}|^\gamma}, \label{eq:gapeq_1}
\eea
 and the equation
 for
  $\Delta (\omega_m)$ remains intact.

\section{The onset temperature for the pairing}
\label{Tp}

 To obtain $T_p$, it is sufficient to consider the linearized gap equation. It is obtained from \eqref{eq:gapeq_1} by setting $\Phi^*$ to be infinitesimally small. Then $\Phi^*(\omega_{m'})$ in the denominators of \eqref{eq:gapeq_1} can be ignored, and the self energy $\Sigma^*(\omega_m)$ can be approximated by its normal state form. The resulting equations are:
\begin{equation}\label{eq:lineargap}
  \begin{aligned}
    \Phi^*(\omega_m)&=g^\gamma \pi T \sum_{m'\neq m}\frac{\Phi^*(\omega_{m'})}{|\omega_{m'}+\Sigma^*(\omega_{m'})|}\frac{1}{|\omega_m-\omega_{m'}|^\gamma},\\
    \Sigma^*(\omega_m)&=g^\gamma \pi T\sum_{m'\neq m}\frac{\sgn(\omega_{m'})}{|\omega_m-\omega_{m'}|^\gamma}.\\
  \end{aligned}
\end{equation}
by power counting, $\Sigma^* (\omega_m) \propto g^\gamma \omega_m^{1-\gamma}$.  Substituting this into the equation for $\Phi$ in (\ref{eq:lineargap}), we obtain that
   the pairing kernel $K_{m,m'} \equiv g^\gamma/(|\omega_{m'}+\Sigma^*(\omega_{m'})|)/|\omega_m -\omega_{m'}|^\gamma$
    is marginal at  $g > |\omega_{m'}| > |\omega_m|$ ($K_{m,m'} \propto
   1/|\omega_m'|$),
   and decays as $K_{m,m'} \propto g^\gamma /|\omega_{m'}|^{1+\gamma}$ at $|\omega_{m'}|> g, \omega_m$.   This implies that $T_p$, if it exists, should be generally of order $g$. The marginal form of the kernel
    is similar to that in the  BCS case,  and within the perturbation theory gives rise to the logarithmical growth of the pairing susceptibility. However, in distinction to BCS,  the marginal form of $K_{m,m'}$   holds only if $|\omega_{m'}| > |\omega_m|$, i.e., at each order of perturbation, the logarithm is cut by the running frequency in the next cross-section in the Cooper ladder.  As the consequence,  the summation of the logarithms alone does not lead to the divergence of the pairing susceptibility~\cite{Wang2016}.
      In this situation, it would be natural to expect  that the pairing  becomes a threshold phenomenon, i.e., it only develops when the effective coupling constant (defined in the next Section) exceeds some finite value.

\begin{figure}
  \begin{center}
    \includegraphics[width=8cm]{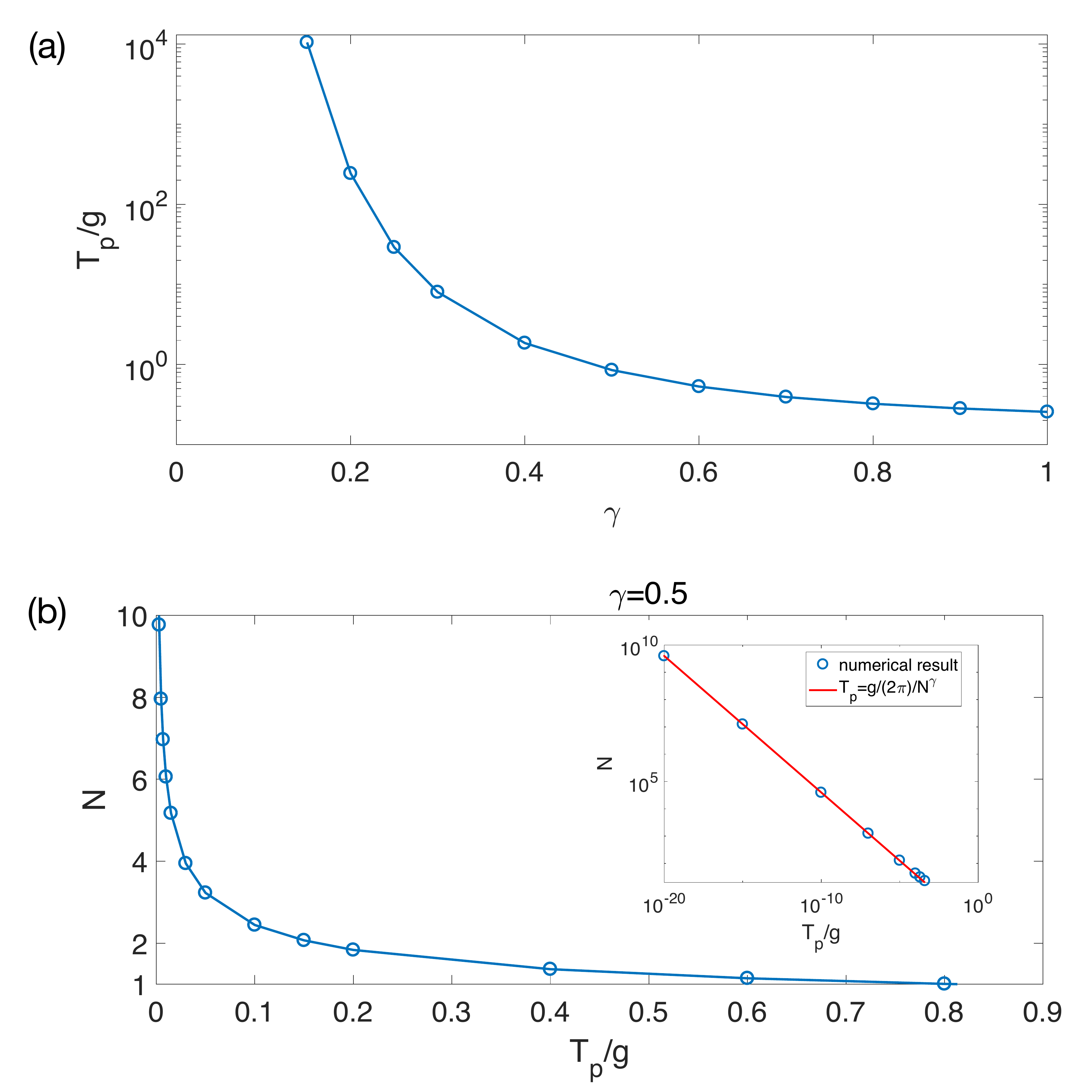}
    \caption{The onset temperature of the pairing, $T_p$,  obtained by solving the linearized equation for the pairing vertex.(a) $T_p$ as a function of $\gamma$ for $N=1$ (the original $\gamma$ model). (b) $T_p$ as a function of $N$  for chosen  $\gamma=0.5$.  The  inset shows a good agreement between the numerical results at large $N$  and the scaling behavior $T_p=\frac{g}{2\pi}\frac{1}{N^{1/\gamma}}$ obtained by considering only the pairing between fermions with first Matsubara frequencies $\omega_m = \pm \pi T$.}\label{fig:Tp}
  \end{center}
\end{figure}

In Fig.\ref{fig:Tp}(a) we show the solution of (\ref{eq:lineargap}). We see that the  onset temperature of the pairing
 $T_p (\gamma)$ is finite for all $\gamma <1$.   The divergence of $T_p (\gamma)$ at vanishing $\gamma$ is just the consequence of the fact that in this limit the interaction decays very slowly ($\gamma =0$ corresponds to BCS limit without upper cutoff). Still, observe that for $\gamma \leq 1$, $T_p \geq g$, i.e., the pairing instability emerges at $T$ above the upper edge of NFL behavior. From this perspective, the pairing does not allow NFL behavior to even develop.

\section{Extension to large $N$}
\label{largeN}

We now analyze whether the existence of a finite $T_p$ is because the tendency to pairing is just numerically stronger than the one to NFL ground state, or there is more fundamental reason why
    the pairing wins.  With this in mind  we extend the $\gamma$ model 
 so as
     to have a parameter
     measuring the
      relative strength of the interaction in particle-particle and particle-hole channels. For this we multiply the coupling in the particle-particle channel by a factor $1/N$, i.e., set it to be $g^\gamma/N$ instead of $g^\gamma$, and keep the coupling in the particle-hole channel intact.
   We will treat $N$ as a free parameter, but keep in mind that in the end we are interested in the system behavior in the physical case of $N =1$. The extension to $N >1$ (albeit a discrete one)  can be formalized if we extend our original model to matrix  $SU(N)$ model\cite{raghu_15, Wang18}.

The modified equations for $\Phi^* (\omega_m)$ and ${\tilde \Sigma}^* (\omega_m)$ are
\bea \label{eq:gapeq_1}
   && \Phi^* (\omega_m) =\nonumber\\
&&
    \frac{\pi T}{N}  g^\gamma \sum_{m' \neq n} \frac{\Phi^* (\omega_{m'})}{\sqrt{({\tilde \Sigma}^* (\omega_{m'}))^2 +(\Phi^* (\omega_{m'}))^2}}
    ~\frac{1}{|\omega_m - \omega_{m'}|^\gamma}, \nonumber \\
   &&  {\tilde \Sigma}^* (\omega_m) = \omega_m+\nonumber\\
&&
     g^\gamma
    \pi T \sum_{m' \neq m}  \frac{{\tilde \Sigma}^* (\omega_{m'})}{\sqrt{({\tilde \Sigma}^* (\omega_{m'}))^2  +(\Phi^* (\omega_{m'}))^2}}
    ~\frac{1}{|\omega_m - \omega_{m'}|^\gamma}, \nonumber
\eea
 and the equation
 for
  $\Delta (\omega_m)$ becomes
 \bea
 &&  \Delta (\omega_m) = \nonumber\\
&&\frac{\pi T}{N}  g^\gamma \sum_{m' \neq m} \frac{\Delta (\omega_{m'}) -N  \Delta (\omega_m) \frac{\omega_{m'}}{\omega_m}}{\sqrt{\omega^2_{m'} +\Delta^2 (\omega_{m'})}}
    ~\frac{1}{|\omega_m - \omega_{m'}|^\gamma}.
     \label{ss_111}
  \eea

Below we will occasionally refer to the equation on $\Phi^* (\omega_m)$ as the gap equation, notwithstanding that the true gap equation is given by Eq. (\ref{ss_111}).  Indeed, once we know $\Phi^* (\omega_m)$ and ${\tilde \Sigma}^* (\omega_m)$, we also know $\Delta^* (\omega_m) = \Phi^* (\omega_m) \omega_m/{\tilde \Sigma}^* (\omega_m)$.

In Fig.\ref{fig:Tp}(b) we show the solution for $T_p (N)$ at a fixed $\gamma =0.5$.
We see that $T_p (N)$ remains finite for {\it all} $N$, i.e., for arbitrary weak strength of the pairing interaction. This result is in clear contradiction with the reasoning above that the pairing at a QCP is a threshold phenomenon.

On a more careful look at the Eliashberg equations we see the reason -- the power-counting argument that $\Sigma^* (\omega_m) \propto \omega^{1-\gamma}_m$ does not work for the first two Matsubara frequencies $\omega_m = \pm \pi T$. For these frequencies,  Eq. (\ref{eq:lineargap}) yields $\Sigma^* (\pm \pi T) =0$
     because contributions from positive and negative $\omega_{m'}$ exactly cancel out.
     To see the consequence of $\Sigma^* (\pm \pi T) =0$,
     consider the equation for $\Phi (\omega_m)$ in the limit $N \gg 1$  and set external $\omega_m = \pi T (2m+1)$ to $\pi T$ (i.e., set $m=0$).
        For $m' = O(1)$, but $m' \neq - 1$, the product $\pi T K_{0,m'}$ is independent of $T$ and is of order $1/N$.  However, for $m' = -1$ ($\omega_{m'} = - \pi T$),
         $\pi T K_{0,-1} = (1/N) (g/(2\pi T))^\gamma$ becomes large at small enough $T$.  A simple experimentation shows \cite{Wang2016} that
          in this situation the Eliashberg equation for $\Phi (\omega_m)$ for different $\omega_m$
           reduces to
     \bea
     &&\Phi^* (\pi T) \approx \frac{1}{N} \left(\frac{g}{2\pi T}\right)^\gamma \Phi^* (-\pi T) \nonumber \\
     &&\Phi^* (\omega_{m>0}) =  \frac{1}{N} \left(\frac{g}{2\pi T}\right)^\gamma \left[\frac{\Phi^* (\pi T)}{|\frac{1}{2}- \frac{\omega_{m'}}{2\pi T}|^\gamma} + \frac{\Phi^* (-\pi T)}{|\frac{1}{2}+ \frac{\omega_{m'}}{2\pi T}|^\gamma}\right].
     \label{s_1}
     \eea
     We will be searching for even-frequency solution $\Phi^* (\omega_m) = \Phi^* (- \omega_m)$.  Then the first equation in (\ref{s_1})  sets $T_p = (g/2\pi) 1/N^{1/\gamma}$, and the second shows that a non-zero $\Phi^* (\omega_m)$ is  induced by $\Phi^* (\pm \pi T)$
     and is suppressed by $N^{1/\gamma}$ for $T\to T_p$.

In the insert for Fig.\ref{fig:Tp}(b) we show the actual $T_p$ vs  $T_p = (g/2\pi) 1/N^{1/\gamma}$. We see that the agreement is perfect at large $N$.

\section{Solution of the full Eliashberg equations below $T_p (N)$}
\label{finiteT_1}

We now study the consequences of the fact that the pairing, at least for large $N$, is fully induced by fermions with Matsubara frequencies $\omega_m = \pm \pi T$. For this we solve non-linear gap equation below $T_p$ and found $\Phi^* (\omega_m)$, ${\tilde \Sigma}^* (\omega_m)$ and $\Delta^* (\omega_m)$. We then use these solutions as inputs and obtain $\Phi^* (\omega)$, ${\tilde \Sigma}^* (\omega)$ and $\Delta^* (\omega)$ in real frequencies.  The full analysis is presented in Ref.\cite{wu_1,abanov_last} Here we briefly describe the main results.

\subsection{Large $N \gg 1$.}
\begin{figure}
  \begin{center}
    \includegraphics[width=7cm]{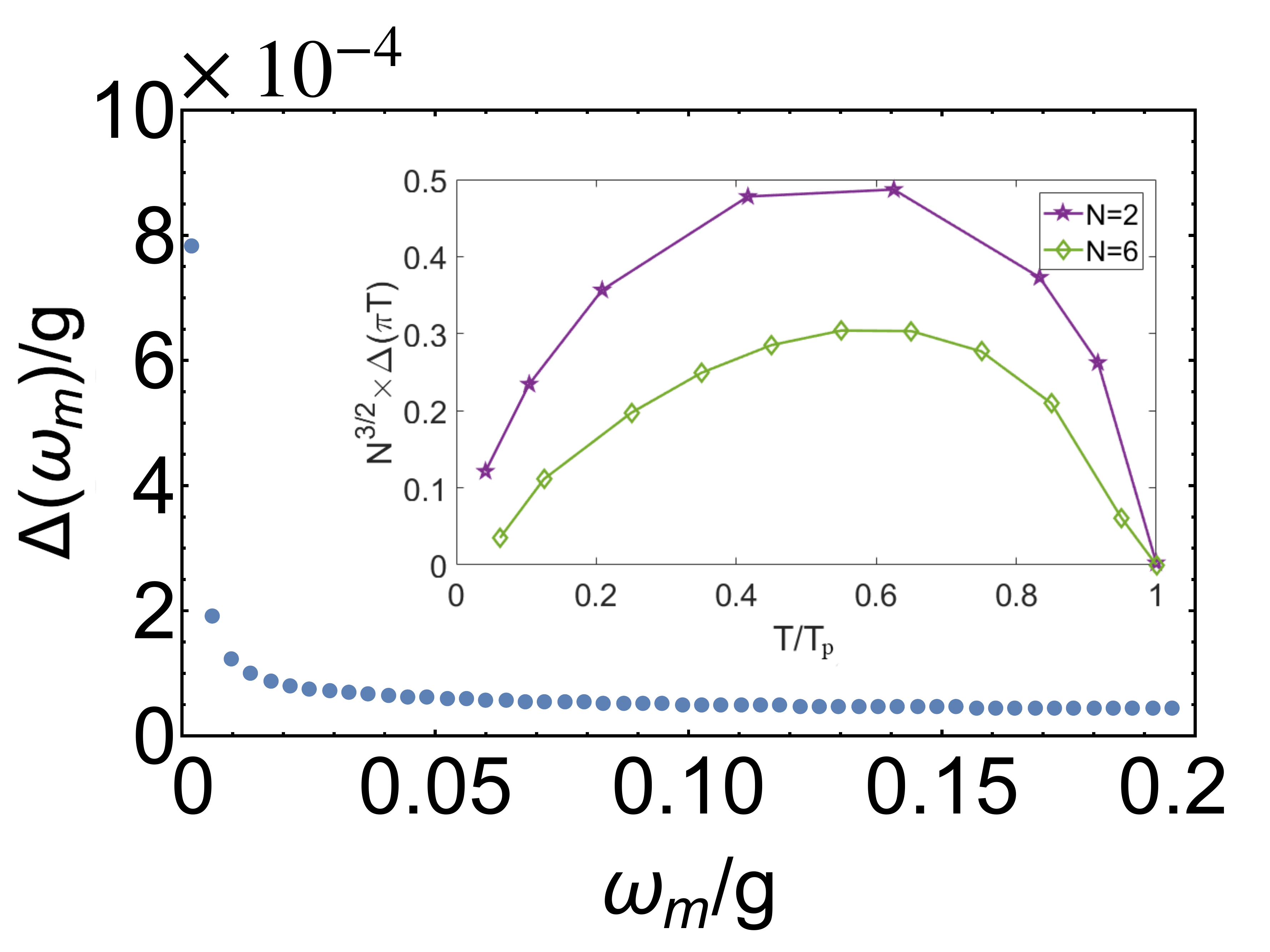}
    \caption{The frequency dependence of the gap function $\Delta(\omega_m)$ at large $N$ ($N > N_{cr}$). nly positive $\omega_m$ are shown. We consider even frequency solution, $\Delta (-\omega_m) = \Delta (\omega_m)$.
      For definiteness we set $\gamma=0.9$, $N=10$ and $T=0.01T_p$. Observe that the value of the gap at first fermionic Matsubara frequency  $\Delta(\pi T)$ is much larger than that at all other $\omega_m$.
       Inset: The temperature dependence of  $\Delta(\pi T)$ for $\gamma=0.9$ and two different $N>N_{cr}$. Observe that the gap is non-monotonic: it emerges at $T = T_p$ and vanishes at $T=0$. The same holds for the gap at all other Matsubara frequencies. }\label{fig:Delta}
  \end{center}
\end{figure}
Because the pairing is induced by fermions with $\omega_m = \pm \pi T$, it is natural to expect that the pairing gap below $T_p$ is much larger at $\omega_m = \pm \pi T$ than at other frequencies. The solution
 of the non-linear Eliashberg equation confirms this: at $\omega_m >0$
 \bea
 &&  \Delta (\pi T) = \pi T \left(\frac{2}{N}\right)^{1/2} \left(1 - \left(\frac{T}{T_p}\right)^\gamma\right)^{1/2} \nonumber \\
&& \Delta (\omega_m > \pi T) =  \frac{1}{N} \frac{\Delta (\pi T)}{H(m,\gamma)}
 \left(\frac{1}{m^\gamma} + \frac{1}{(m+1)^\gamma}\right) \propto\nonumber\\
&&
  T \left(\frac{2}{N}\right)^{3/2} \left(1 - \left(\frac{T}{T_p}\right)^\gamma\right)^{1/2}.
\label{b_4}
\eea
 where  $H(a,b) = \sum_{1}^a n^{-b}$ is a Harmonic number. We plot $\Delta (\omega_m)$ in Fig.\ref{fig:Delta}
 We also see from (\ref{b_4}) that $\Delta (\omega_m)$ vanishes at $T=0$, i.e., the normal state is a naked NFL, although at the verge of a pairing instability. We show $\Delta (\pi T)$ vs $T$ in the insert to Fig.\ref{fig:Delta}  This reentrant behavior of the gap is the direct consequence of the pairing induced by fermions with $\omega_m = \pm \pi T$ as at $T=0$ a Matsubara frequency becomes a continuous variable and fermions with $\omega_m = \pm \pi T$  cannot play any special role.

\begin{figure*}
   \begin{center}
     \includegraphics[width=17cm]{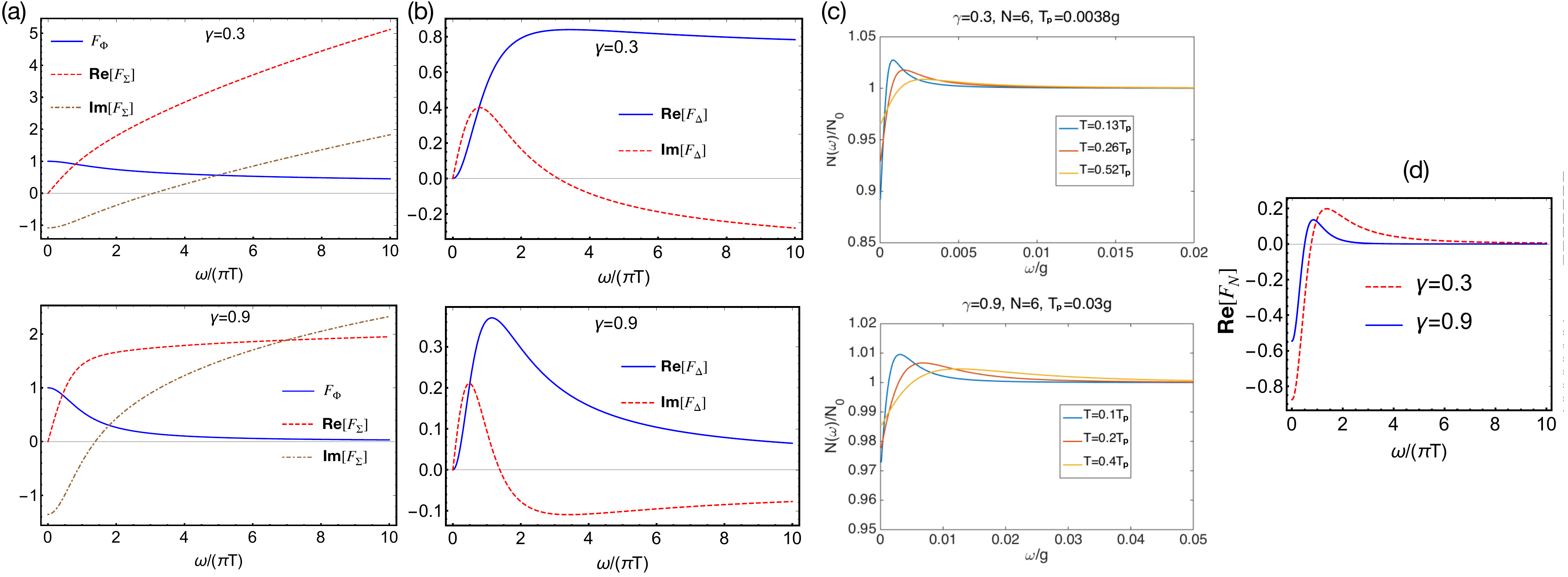}
     \caption{(a)-(b) The scaling functions $F_{\Phi}(\frac{\omega}{\pi T})$ for the pairing vertex, $F_{\Sigma}(\frac{\omega}{\pi T})$ for the fermionic self-energy, and $F_{\Delta}(\frac{\omega}{\pi T})$ for the gap function (see Eq. (\ref{s_20})) for representative  $\gamma=0.3$ and $\gamma=0.9$; (c) The DOS $N(\omega)$ for the same $\gamma$ and $N=6$. The DOS have been obtained by solving Eliashberg equations on the real axis, using the solution on Matsubata axis as an input.   (d) The scaling function for the DOS,  $\Ree[F_N(\frac{\omega}{\pi T})]$, see Eq. (\ref{s_21}).}\label{fig:scaling}
   \end{center}
 \end{figure*}

 Using the solution along the Matsubara axis, one can obtain the gap function along the real axis $\Delta (\omega)$.  This requires one to solve the set of integral equations for $\Phi^* (\omega)$ and ${\tilde \Sigma}^* (\omega)$ with the solution along the  Matsubara axis as an input.  We skip the details (see Ref\cite{wu_1}) and present the results.  The vertex function $\Phi^* (\omega)$, the self-energy $\Sigma^* (\omega)$, and the gap function $\Delta (\omega)$ are given by
 \bea
 && \Phi^* (\omega) =  \left(\frac{2}{N}\right)^{3/2} \pi T  \left(\frac{g}{\pi T}\right)^\gamma  \left(1 - \left(\frac{T}{T_p}\right)^\gamma\right)^{1/2}
  F_\Phi \left(\frac{\omega}{\pi T}\right), \nonumber \\
&& \Sigma^* (\omega)  = \pi T  \left(\frac{g}{\pi T}\right)^\gamma  F_\Sigma \left(\frac{\omega}{\pi T}\right), \nonumber \\
&& \Delta(\omega)  =  \left(\frac{2 }{N}\right)^{3/2} \pi T \left(1 - \left(\frac{T}{T_p}\right)^\gamma\right)^{1/2}  F_\Delta \left(\frac{\omega}{\pi T}\right),
\label{s_20}
\eea
where $F_\Phi, F_\Sigma$ and $F_\Delta$ are scaling functions of $\omega/\pi T$. We remind that ${\tilde \Sigma}^* (\omega) = \omega + \Sigma^* (\omega)$.  We plot
 these functions in Fig.\ref{fig:scaling}  Because $\Phi^* (\omega) \propto 1/N^{3/2}$ is small, the self-energy
 in (\ref{s_20}) retains, to order $1/N^3$,  the same NFL form as in the normal state, i.e., there is essentially no feedback effect on fermions from the pairing. At large argument, $F_\Sigma (x) \propto x^{1-\gamma}e^{i\pi\gamma/2}$, i.e., $\Sigma^* (\omega) \propto \omega^{1-\gamma}$. We also note that
 at small $\omega$, Re $\Delta (\omega) \propto \omega^2$ and Im $\Delta (\omega) \propto \omega$.  This behavior is a signature of a gapless SC.
The DOS is
\bea
&& N(\omega) =  N_0 \Ree  \left[\frac{\omega}{(\omega^2 - \Delta^2 (\omega))^{1/2}}\right] \label{s_21} \\
&& \approx N_0 \left(1 + \frac{1}{2}\left(\frac{2}{N}\right)^{3} \left(1 - \left(\frac{T}{T_p}\right)^\gamma\right) \Ree  \left[F_N \left(\frac{\omega}{\pi T}\right)\right]\right). \nonumber
\eea
where $N_0$ is the DOS in the normal state.
We see that the magnitude of
$N(\omega)/N_0 -1$
 depends on $T/T_p$. However, the frequency dependence of the DOS is determined by $F_N (\omega/(\pi T))$, which for any given $\gamma$ is a universal function of $\omega/T$  and  does not depend on $T/T_p$.  This implies that the characteristic frequency, at which $N(\omega)$ deviates from $N_0$, is determined by the temperature rather than by the magnitude of the superconducting gap. We show the DOS in Fig.\ref{fig:scaling}(d)

\subsection{Smaller $N \geq 1$.}\label{sec:smallerN}
\begin{figure*}
  \begin{center}
    \includegraphics[width=17cm]{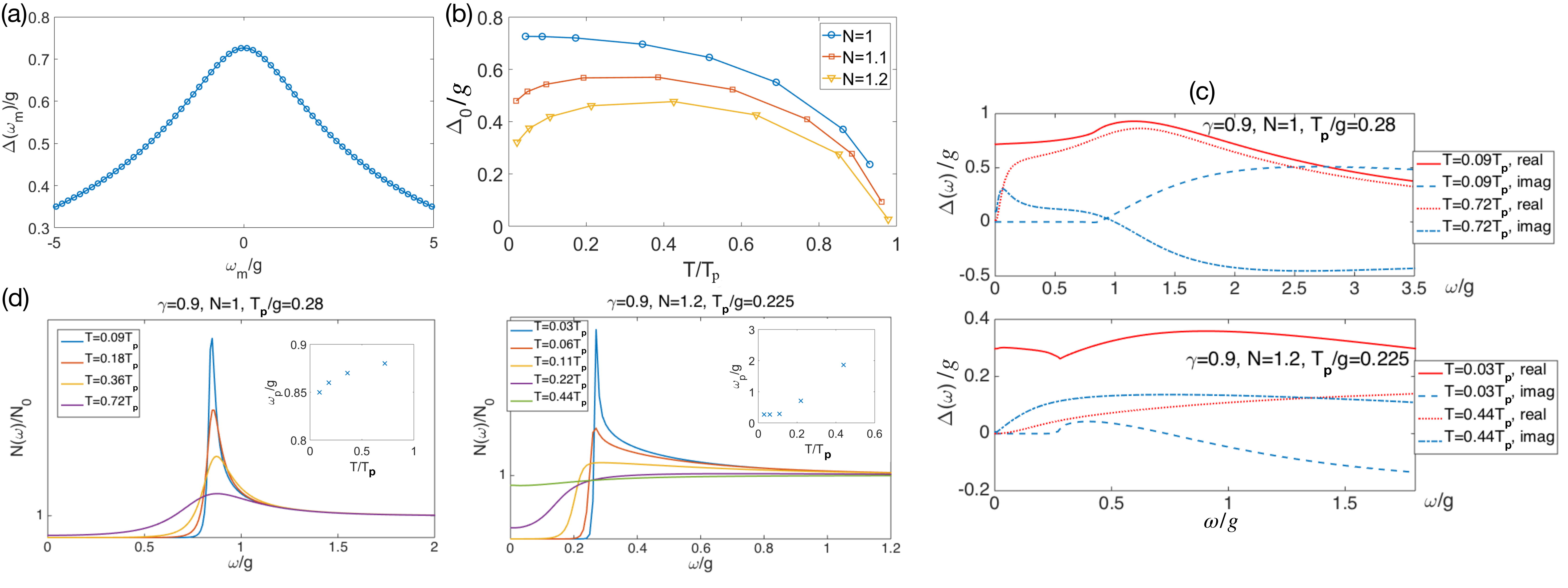}
    \caption{Smaller $N < N_{cr}$. (a) The gap function $\Delta(\omega_m)$ along Matsubara frequency axis. We set $\gamma=0.9$, $N=1$ and $T=0.18T_p$. Observe that $\Delta (\omega_m)$ is no longer strongly peaked at $\omega = \pm \pi T$; (b) The gap function at $\omega_m = \pi T$  as a function of temperature for $\gamma=0.9$. The gap now tends to a finite value at $T \to 0$; (c) Real and imaginary parts of the gap function $\Delta(\omega)$ along real frequency axis for $T<T_{\crr}$ and $T>T_{\crr}$. At  $T< T_{\crr}$ the gap function resembles that of a non-critical BCS/Eliashberg superconductor, i.e. at small $\omega$ it is real and weakly dependent on $\omega$.  At higher $T < T_{\crr}$, the functional form of $\Delta (\omega)$  is  similar to the one obtained in at $N > N_{cr}$, see Fig.\ref{fig:scaling}. (d) The DOS $N(\omega)$ for various $T$. At low $T<T_{\crr}$ the DOS has a sharp peak at $\omega=\Delta_0$ and nearly vanishes for $\omega<\Delta_0$. At higher $T>T_{\crr}$ the behavior of the DOS is similar to the one at $N> N_{cr}$, i.e.,  the position of the maximum of $N(\omega)$ shifts to a higher frequency with increasing temperature. The insets show the position of the maximum, $\omega_p$, as a function of $T/T_p$.}\label{fig:smallN}
  \end{center}
\end{figure*}
To avoid a lengthy discussion, here we present only the numerical results. In Fig.\ref{fig:smallN}(a) we show
 the gap function  along the Matsubara axis. We see that now $\Delta (\omega_m)$ is a smooth function of frequency, i.e., fermions with $\omega_m = \pm \pi T$ are no longer special. In Fig.\ref{fig:smallN}(b) we show
 $\Delta (\pi T)$ as a function of $T/T_p$. We clearly see that the gap now reaches a finite value $\Delta$  at $T=0$.
This implies that the ground state is now a superconductor. We show $\Phi^* (\omega)$, $\Sigma^* (\omega)$ and $\Delta (\omega)$ in Fig.\ref{fig:smallN}(c) and the DOS in Fig.\ref{fig:smallN}(d) At low $T$ the system now displays BCS-type behavior. Namely, $\Sigma^* (\omega)$ acquires a FL form due to feedback from the pairing ($\Sigma^* (\omega)$ is linear in $\omega$ at small frequencies),
    the gap $\Delta (\omega)$ is predominantly real and reaches $\Delta$  at $\omega =0$, and
       DOS has a sharp peak at $\omega = \Delta$, which initially moves to a smaller $\omega$ as $T$ increases, consistent with the "gap closing" behavior.

 At higher $T$, above some $T_{\crr} < T_p$,  the  system behavior changes -- the self-energy recovers its NFL, normal state form, the gap function becomes predominantly imaginary at small $\omega$, and the DOS $N(\omega)$ displays $\omega/T$ scaling, instead of $\omega/\Delta$ one, and shows "gap filling" behavior.
 This is the same behavior that we found at large $N$.

\begin{figure}
   \begin{center}
     \includegraphics[width=7cm]{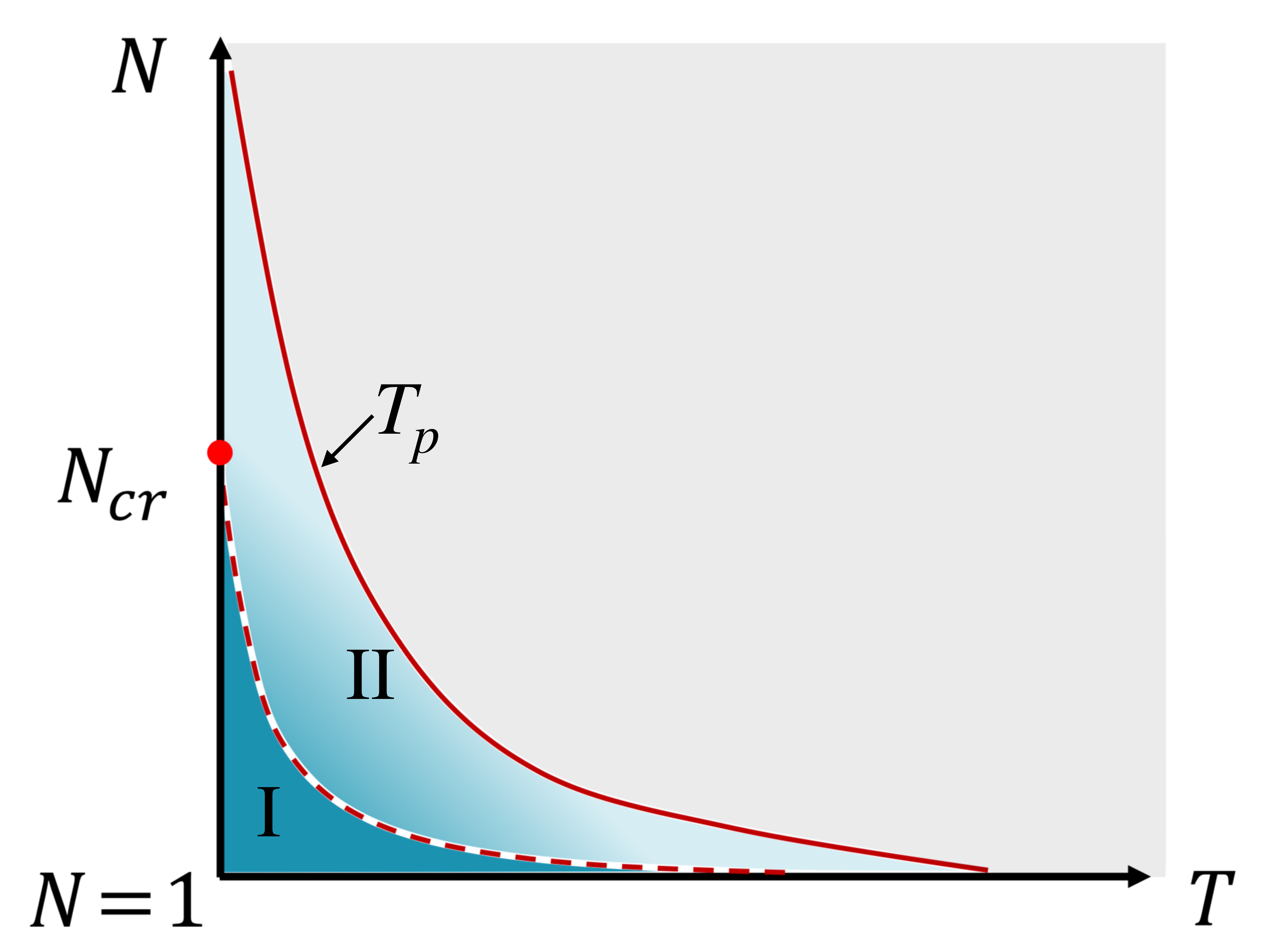}
     \caption{The phase diagram constructed from the numerical results in Sec.\ref{finiteT_1}. The red solid line is $T_p(N)$. The dashed red line makes the crossover between BCS-like "gap closing" behavior  in region I and non-BCS "gap filling" behavior in region II.}\label{fig:phasediagram}
   \end{center}
 \end{figure}

 These results show that at $N \ge1$, the system undergoes a crossover between BCS-like behavior at small $T$ and non-BCS, "gap filling" behavior at higher $T$.  In Fig.\ref{fig:phasediagram} we show the phase diagram, extracted from the numerical data.  For a given $\gamma$, the  crossover temperature $T_{\crr}$ is finite for $N =1$, gets smaller with increasing $N$, and vanishes at some $N_{cr}$, whose value depends on $\gamma$.
 In the rest of this paper we analyze what determines the crossover line $T_{\crr} (N)$ and why it ends up at a finite $N= N_{cr}$.  For this we first consider the case $T=0$.

\section{Solution of Eliashberg equations at $T=0$}
\label{T=0}

In this section we present the reasoning for the existence of critical $N= N_{cr}$, separating NFL ground state at $N > N_{cr}$ and a SC state at $N < N_{cr}$.  We present semi-qualitative reasoning, which works best at small $\gamma$. For rigorous consideration see Ref.\cite{ac_exact}.

At $T=0$ the Eliashberg equations for $\Phi^* = \Phi$ and ${\tilde \Sigma}^* = {\tilde \Sigma}$ are
 \bea
   && \Phi (\omega_m) =\nonumber\\
&&
    \frac{g^\gamma}{2N}  \int d \omega'_m  \frac{\Phi (\omega'_{m})}{\sqrt{{\tilde \Sigma}^2 (\omega'_{m})) +\Phi^2 (\omega'_{m}))}}
    ~\frac{1}{|\omega_m - \omega'_{m}|^\gamma}, \label{eq:gapeq_1} \\
   &&  {\tilde \Sigma} (\omega_m) = \omega_m\nonumber\\
&&
   +  \frac{g^\gamma}{2} \int d \omega'_m
      \frac{{\tilde \Sigma} (\omega'_m)}{\sqrt{{\tilde \Sigma}^2 (\omega'_{m})  +\Phi^2 (\omega'_{m})}}
    ~\frac{1}{|\omega_m - \omega'_{m}|^\gamma}, \label{eq:gapeq_1_1}
\eea
 In the normal state,
  \beq
  {\tilde \Sigma} (\omega_m) = \omega_m + \omega_0^\gamma |\omega_m|^{1-\gamma} \sgn{\omega_m}
\label{ss_11_1}
\eeq
 where $\omega_0 = g/(1-\gamma)^{1/\gamma}$.   At small $\gamma$, $\omega_0 = g/e$.
 In the limit of  infinitesimally small $\Phi (\omega_m)$ we obtain, using (\ref{ss_11_1})
  \beq
      \Phi (\omega_m) =\frac{1-\gamma}{2N} \int d \omega'_m \frac{\Phi(\omega'_m)}{|\omega'_{m}| ^{1-\gamma} |\omega_m-\omega'_{m}|^\gamma}
 \frac{1}{1 + \left(\frac{|\omega'_m|}{\omega_0}\right)^\gamma}
  \label{eq:lineargap_1}
  \eeq
 After rescaling ${\bar \omega}_m = \omega_m/\omega_0$, this equation
 becomes completely universal, with $N$ as the only parameter:
   \beq
      \Phi ({\bar \omega}_m) =\frac{1-\gamma}{2N} \int d {\bar \omega}'_m \frac{\Phi({\bar \omega}'_m)}{|{\bar \omega}'_{m}| ^{1-\gamma} |{\bar \omega}_m-{\bar \omega}'_{m}|^\gamma}
 \frac{1}{1 + |{\bar \omega}'_m|^\gamma}
  \label{eq:lineargap_1_1}
  \eeq

\subsection{Large $N$}

Here we consider large $N$.
 The effective coupling constant in (\ref{eq:lineargap_1}) scales as $1/N$, hence the solution with a non-zero $\Phi (\omega_m)$ emerges only if the smallness of the coupling is compensated by a large value of the frequency integral in the r.h.s. of (\ref{eq:lineargap_1_1}).  This is indeed what happens in a BCS superconductor (the case $\gamma =0$), where  the pairing kernel  scales as $1/|\omega_m|$,  $\Phi (\omega)= \Phi$ is independent on the running fermionic frequency,
  and the integral $\int_0^{\omega_D} d \omega_m  \Phi/|\omega_m|$, with some upper cutoff at $\omega_D$, is logarithmically singular. This gives rise to a divergence of $\Phi$ at some non-zero total incoming frequency $\Omega_{tot}$.

   For a non-zero $\gamma$,  the pairing kernel is the function of both internal $\omega'_m$ and external $\omega_m$:
\beq
 K(\omega_,\omega'_m) = \frac{1}{|\omega'_{m}|^{1-\gamma} |\omega_m-\omega'_{m}|^\gamma~ (1 + \left(\frac{|\omega'_m|}{\omega_0}\right)^\gamma)},
 \eeq
 If we set the external $\omega_m$ to zero, we find that $K (0,\omega'_m) =\left[ |\omega'_{m}| (1 + |\omega'_m|/\omega_0)^\gamma)\right]^{-1}$ is marginal at $|\omega'_m| <\omega_0$,  like in BCS theory.  This implies that if we again add $\Phi_0$ and compute $\Phi (\Omega_m)$ perturbatively, the series are logarithmical, like in BCS case.  In distinction to BCS, however, each logaritmical integral $\int d \omega'_m/|\omega'_m|$  runs between $|\omega'_m| \sim \omega_0$, which sets the upper limit, and
   $|\omega'_m| \sim |\omega_m|$, which sets the lower limit. We can then safely set $\Omega_{tot} =0$.
   Summing up logarithmical series we then obtain
\bea
\Phi (\omega_m) &=& \Phi_0 \sum_{k=0}^{\infty }\frac{1}{k!}\left[\frac{1-\gamma }{N}\log \frac{\omega_{0}}{|\omega_{m}|} \right] =\Phi_0 \left[\frac{\omega_0}{|\omega_m|}\right]^{\frac{1-\gamma}{N}}
\label{su_2}
\eea
We see that $\Phi (\omega_m)$ does not diverge at any non-zero $\omega_m$.  The implication is that, at a finite $\gamma$, summation of the  logarithms does not give rise to pairing instability.

We now go beyond perturbation theory and analyze the linearized equation for $\Phi (\omega_m)$, Eq. (\ref{eq:lineargap_1}), without the $\Phi_0$ term.  Our first observation is that the power-law solution $\Phi (\omega_m) \propto \left(\omega_0/|\omega_m|\right)^{(1-\gamma)/N}$, which we found by summing up logarithms, does satisfy Eq.  (\ref{eq:lineargap_1}) at small frequencies $|\omega_m| \ll \omega_0$,  when one can neglect $(|\omega_m|/\omega_0)^\gamma$ term in the denominator in  (\ref{eq:lineargap_1}).  To see this, we note that  $\Phi (\omega_m) \propto \left(\omega_0/|\omega_m|\right)^{(1-\gamma)/N}$  does satisfy the truncated version of Eq. (\ref{eq:lineargap_1}) if
\beq
1 = \frac{(1-\gamma)}{2N} \int \frac{dx}{|x|^{(1-\gamma)(N+1)/N}} \frac{1}{|1-x|^\gamma}
 \label{nn_1}
 \eeq
One can verify that this condition is satisfied to order $O(1/N)$ -- the compensating factor $N$  comes from large $|x| \gg 1$  in the integral.

 We now argue that there is another  possibility to compensate for the $1/N$ smallness of the coupling constant in (\ref{eq:lineargap_1}), by choosing
$\Phi (\omega_m) \propto \left(\omega_0/|\omega_m|\right)^{\gamma -(1-\gamma)/N}$, such that the integral over $\omega'_m$ in the r.h.s. of (\ref{eq:lineargap_1}) almost diverges at small $\omega'$. Indeed, substituting this form into the truncated version of (\ref{eq:lineargap_1}) and rescaling, we find that the equation is satisfied if
\beq
1 = \frac{(1-\gamma)}{2N} \int \frac{dx}{|x|^{1-(1-\gamma)/N}} \frac{1}{|1-x|^\gamma}
 \label{nn_2}
 \eeq
One can verify that this condition is again satisfied to order $O(1/N)$ -- the compensating factor $N$  now comes from small $|x| \ll 1$  in the integral. Note that this solution could not be obtained within a conventional logarithmic approximation (or, equivalently,  RG scheme) as the latter assumes that  the logarithms, which sum up into anomalous power-law form,  come from internal frequencies larger than the external one.

The full solution  for $\Phi (\omega_m)$  at $|\omega_m| \ll \omega_0$ is the combination of the two power-law forms:
\beq
 \Phi (\omega_m) =  \frac{C_1}{|\omega_m|^{\gamma(1/2-b)}} + \frac{C_2}{|\omega_m|^{\gamma (1/2 + b)}}
\label{nn_3}
\eeq
where at large $N$,
 $b^{2} \approx 1/4-(1-\gamma)/(N\gamma)$.
The overall factor doesn't matter because $\Phi (\omega_m)$ is defined up to a constant multiplier, but the ratio $C_2/C_1$ is a free parameter at this moment.

We now verify whether by properly choosing
$C_2/C_1$ one can extend the solution to larger $\omega_m$, when $(|\omega_m|/\omega_0)^\gamma$ term in the denominator of (\ref{eq:lineargap_1}) cannot be neglected.    For this we fist note that at large $|\omega_m| \gg  \omega_0$,
 $\Phi (\omega_m) \propto 1/|\omega_m|^\gamma$ because in this limit the external $\omega_m$ can be pulled  out from the integral in the r.h.s. of (\ref{eq:lineargap_1}) and the remaining integral converges at $|\omega'_m| = O(\omega_0)$.
 To study the crossover from small to large frequencies we consider $\gamma \ll 1$.

For these $\gamma$,
   the compensation of $1/N$ in the integral in the r.h.s. of (\ref{eq:lineargap_1}) comes from internal $\omega'_m$ either much larger or much smaller than external $\omega_m$. Accordingly, we split the integral over $\omega'_m$ into two contributions and  approximate $|\omega_m - \omega'_m|$ by $|\omega'_m|$ in the one and   by $|\omega_m|$ in the other. A simple experimentation and rescaling shows that the integral equation for the pairing vertex then reduces to
    \beq
    \Phi (x) = b \left[\int^{\infty}_{x} dy \frac{\Phi (y)}{y (1+y)} + \frac{1}{x} \int_0^x dy \frac{\Phi (y)}{1+y}\right]
   \label{su_6}
   \eeq
   where we introduced $x = (|\omega|/\omega_0)^\gamma$.
Differentiating twice over $x$, we obtain second order differential equation
   \beq
   (\Phi (x) x)^{''}  = - (1/4-b^{2}) \frac{\Phi (x)}{(x (x+1)},
   \label{su_7}
   \eeq
   where  $(...)^{''} = d^2 (...)/dx^2$.
   The solution of (\ref{su_7}) is a linear combination of the two hypergeometric functions:
     \begin{widetext}
     \bea
&&\Phi (x) = \frac{1+x}{x^{1/2}} \left(\frac{C_1}{x^{b}} {_2}F{_1} \left[1/2-b, 3/2 -b, 1- 2b; -x\right] +  \frac{C_2}{x^{-b}} {_2}F{_1} \left[1/2+b, 3/2 + b, 1+ 2b; -x\right]\right)
 \label{su_9}
 \eea
 \end{widetext}
where, we remind, $x = (|\omega_m|/\omega_0)^\gamma$.  At small $x$ this reproduces the power-law form of Eq. (\ref{nn_3}). At large $x$ we should have $\Phi (x) \propto 1/x$. Using the asymptotic form of the Hypergeometric function, we obtain from (\ref{su_9}),  $\Phi (x) = A_1/x + A_2$, where $A_1$ and $A_2$ are linear combinations of $C_1$ and $C_2$. To match with high-frequency behavior we need to set $A_2=0$. This determines
the ratio $C_2/C_1$.  For this given $C_2/C_1$, $\Phi (x)$ in (\ref{su_9}) is the true solution of the linearized gap equation, which smoothly interpolates between the two limits. We emphasize that one need to
 fix just one free parameter to obtain the analytic solution of the original {\it integral} equation. This would not be possible if one would artificially set the upper cutoff in (\ref{eq:lineargap_1_1}) at some $x_0$ and use (\ref{nn_3}) for $x < x_0$.  Then one had to satisfy an infinite number of boundary conditions on  $\Phi (x)$ and its derivatives at $x=x_0$, which would be impossible as $C_2/C_1$ is the only parameter.

  We next analyze whether there exists a  solution with a finite
 (i.e., not infinitesimally small) $\Phi (\omega_m)$ and, hence, a finite condensation energy. A way to check this is to take the solution of the linearized gap equation at some large $N = N_0$ as an input, reduce $N$ a bit (i.e., increase the interaction in the particle-particle channel) and check whether there appears a finite $\Phi (\omega_m)$.  We argue that this does not happen because a finite  $\Phi$  would give rise to a divergent  condensation energy $E_c = F_{sc} - F_n$. Indeed, using the Eliashberg formula for the Free energy for the $\gamma$-model~\cite{wu_1,emil}
  and expanding it  in powers of $\Phi$, we find
  \beq
  E_c = D (N_0 - N) \int d \omega_m \frac{\Phi^2 (\omega_m)}{|\omega_m|^{1 - \gamma}} + O(\Phi^4)
  \label{suu_1}
  \eeq
  where $D$ is a numerical prefactor.
     Substituting the small-frequency form of $\Phi (\omega_m)$ from (\ref{nn_3})
       we find that  the  $C_2$ term gives infra-red divergent contribution to the integral in (\ref{suu_1}) in the form
  $\int d \omega_m/|\omega_m|^{1 + b}$.
   The only option to avoid the divergence is to set $C_2 =0$. However, then one would not be able to match low-frequency and high-frequency behavior of $\Phi (\omega_m)$.  The same result is obtained if we directly solve the non-linear gap equation using the solution of the  linearized gap equation as the source -- the frequency integral in the source term  diverges if we keep $C_2$ finite. This implies that that the solution for $\Phi$ in (\ref{su_9}) is not normalizable and only holds if $\Phi (\omega)$ is infinitesimally small.

 We see therefore that for large $N$ the system at $T=0$ is "frozen" at the transition towards pairing:
   the solution of the linearized equation for $\Phi$ exists, but  the non-linear equation has no solutions. This is fully consistent with our analysis in the previous section, where we found that, at large $N$, the pairing gap vanishes at $T=0$.

  We next observe that the above analysis is valid as long as
$b^{2} =1/4-1/(N\gamma) >0$
($b$
in (\ref{nn_3}) and (\ref{su_9}) is real),  i.e., as long as $N > N_{cr} = 4/\gamma$. For smaller $N$ the analysis has to be done differently.

 \subsection{$N = N_{cr}$}

   At $N= N_{cr}$ the two exponents
$\gamma (1/2 \pm b)$
in (\ref{nn_3}) merge into the single one, equal to $\gamma/2$.  At a first glance, this implies that there is no parameter  analogous to $C_2/C_1$, which could be adjusted to match $\Phi (\omega_m)$ at large $\omega_m$. On a more careful look, however, we find that there are fact two solutions at small $x$, when
$b =0$
  \beq
\Phi (\omega_m) =  \frac{1}{|\omega_m|^{\gamma/2}} \left(C_1 + C_2 \log{|\omega_m|} \right)
\label{nn_3_1}
\eeq
\begin{figure}
  \begin{center}
    \includegraphics[width=8cm]{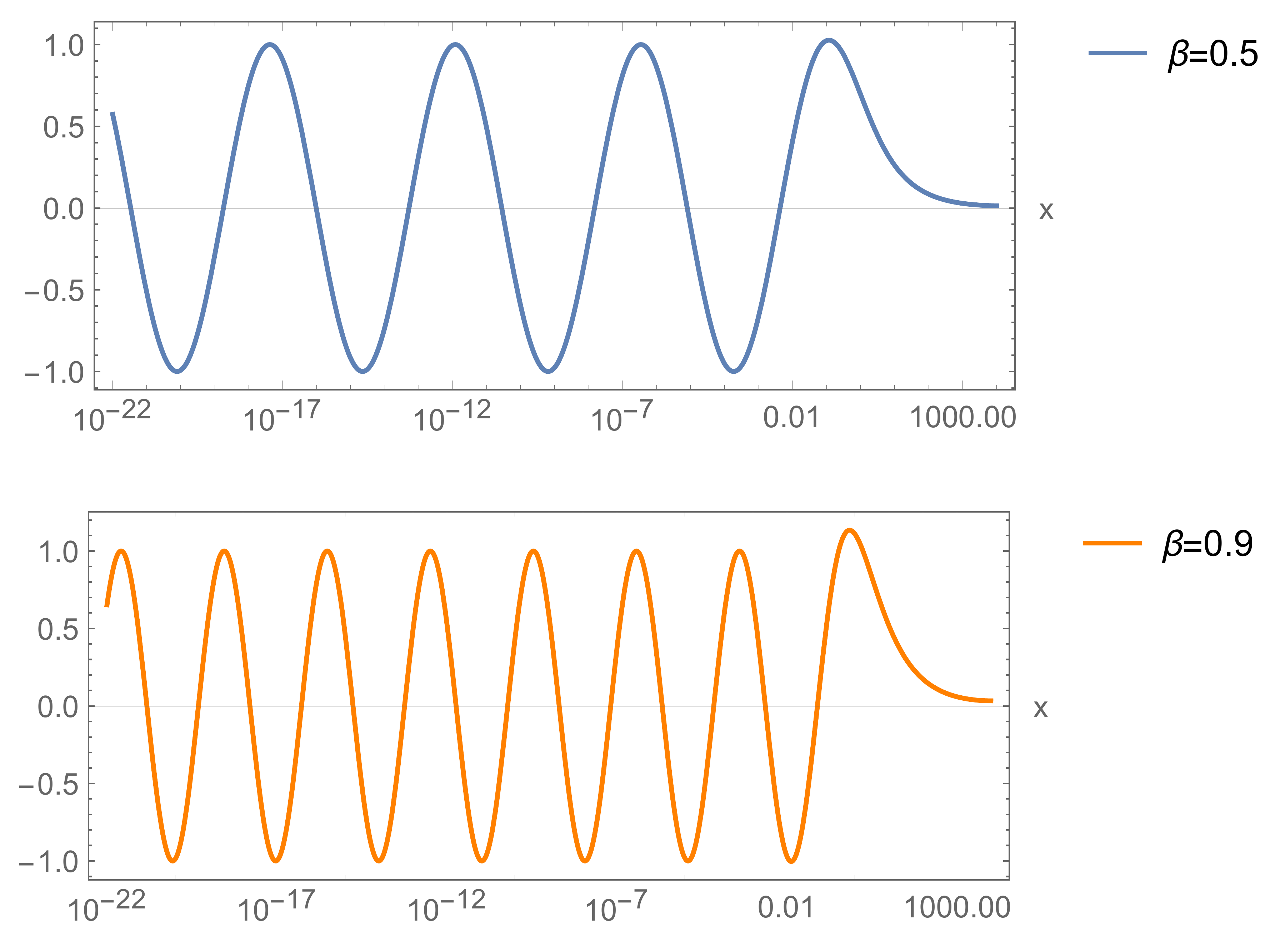}
    \caption{The solution of Eq. \eqref{su_17} for the pairing gap in the case $N < N_{cr}$. The horizontal axis is $x = (\omega/\omega_0)^{\gamma}$, the vertical axis is $x^{1/2}\Phi(x)$.
      The plots are for two different $\beta \propto (N_{cr}-N)^{1/2}$.  Observe that the gap function oscillates at small frequencies and vanishes at high frequencies. Oscillations are on logarithmical scale, and to clearly see them one needs to go to truly small $x$.}\label{fig:Phix}
  \end{center}
\end{figure}
The full solution for $b=0$ is expressed via Hypergeometric and MejerG functions. Like for larger $N$, one can interpolate smoothly between small and large $x$ limits   by adjusting
$C_2/C_1$ ratio.  There is no solution of the non-linear equation.

\subsection{$N < N_{cr}$}

Consider first the linearized equation for the pairing vertex, Eq. (\ref{eq:lineargap_1}). Let's focus on small $\omega_m$, neglect $\omega_m$ compared to the self-energy $\Sigma (\omega_m)$ (i.e., neglect the last term in (\ref{eq:lineargap})) and
 search for the same  power-law solutions
$\Phi (\omega_m) \propto 1/|\omega_m|^{\gamma(1/2 \pm b)}$
as before. Now
$b^{2} = (1-N_{cr}/N)/4 <0$,
i.e., the two exponents are complex conjugated~\cite{acf,raghu_15,Wang2016}. Substituting into (\ref{eq:lineargap}) we find that the solution with the complex exponents exists, despite that all coefficients in (\ref{eq:lineargap}) are real numbers.  It is convenient to define the exponents as $\gamma(1/2 \pm  i \beta)$ where now $2\beta = \sqrt{N_{cr}/N -1} >0$.  Then the power-law solution is, in terms of dimensionless $x = (|\omega_m|/\omega_0)^\gamma$:
\beq
\Phi (x)   =  \frac{C}{x^{1/2}} \cos\left(\beta \log{x} + \phi\right)
\label{nn_3_c}
\eeq
 where  ${C}$ is an irrelevant overall factor. The role formerly played by $C_2/C_1$ is now played by  a phase factor $\phi$, which at this stage is a free parameter.  This $\Phi (x)$ is now oscillating on a logarithmical scale down to the lowest $x$, i.e., the lowest $\omega$.
We note in passing that complex exponents have been detected in other sets of problems, including holographic description of Fermi surfaces~\cite{Liu,Faulkner} and recent studies of scaling dimensions of operators in interacting SYK-type models~\cite{klebanov}.

At large $|\omega_m| \gg \omega_0$, i.e., at $x \gg 1$,  we still can pull $\omega_m$ from the integral in the r.h.s. of  Eq. (\ref{eq:lineargap_1}) and obtain
$\Phi (\omega_m) \propto 1/|\omega_m|^\gamma$, i.e., $\Phi (x) \propto 1/x$.  Like before, we need to verify whether this behavior and the one at small $|\omega_m| \ll \omega_0$ can be matched by choosing a proper $\phi$  in (\ref{nn_3_c}).
For this we again assume that $\gamma$ is small and keep in the integral over $\omega'_m$ in
(\ref{eq:lineargap_1}) the contributions from
   $\omega_m \gg \omega'_m$ and $\omega_m \ll \omega'_m$, and reduce integral equation for $\Phi (x)$ to the same differential equation as in  (\ref{su_7}). Solving this equation for $N < N_{cr}$, we  obtain
\begin{widetext}
     \beq
 \Phi (x) = {\bar C} \frac{1+x}{\sqrt{x}} \times Re\left(e^{-i \phi} x^{i \beta } {_2}F{_1} \left[\frac{1}{2} + i\beta , \frac{3}{2} + i\beta , 1 + 2i\beta ; -x\right] \right)
 \label{su_17}
 \eeq
 \end{widetext}
 where ${\bar C} \sim C$. We plot $\Phi (x)$  in Fig.\ref{fig:Phix}. At $x \ll 1$, this $\Phi (x)$ reduces to the one in (\ref{nn_3_c}).
 At $x \gg 1$,  solution can be expressed in terms of Bessel and Neumann functions as
 \beq
 \Phi (x) = \frac{C}{\sqrt{x}} \left[ a_J J_{1}\left(\sqrt{\frac{N_{cr}}{N x}}\right) + a_Y Y_1 \left(\sqrt{\frac{N_{cr}}{N x}}\right)\right]
 \label{l_1}
 \eeq
 where the $a_J, a_Y$ are expressed in terms of the phase factor $\phi$  in (\ref{su_17}). Using that
 $J_1 (z \ll 1) \sim z$ and $Y_1 (z \ll 1) \propto 1/z$, we find that the required form $\Phi (x) \propto 1/x$ at large $x$ is reproduced if we choose the phase such that $a_Y=0$.

 This consideration shows that the solution of the linearized equation for the pairing vertex exists also for all $N < N_{cr}$.  Combined with earlier analysis, we see that it exists for all values of $N$, including physical $N =1$.  There is however, an essential difference between the form of $\Phi (\omega_m)$: at $N > N_{cr}$ it is a sign-preserving function of $\omega$, while at $N < N_{cr}$ it oscillates down to the lowest $\omega$.

 We now argue that there is a crucial difference between the cases $N < N_{cr}$ and $N = N_{cr}$. Namely,
  for $N < N_{cr}$, the quadratic in $\Phi$ term in the Free energy does not diverge. Indeed, in logarithmical variables the integral in (\ref{suu_1}) now reduces to $\int^1_{-\infty} d y \cos^2 (\beta y + \phi)$. This integral converges at the lower limit if we add infinitesimally small  damping term to the argument of $\cos$.  Because of convergence, the solutions of the non-linear gap equation are now possible.

\begin{figure}
  \begin{center}
       \includegraphics[width=7cm]{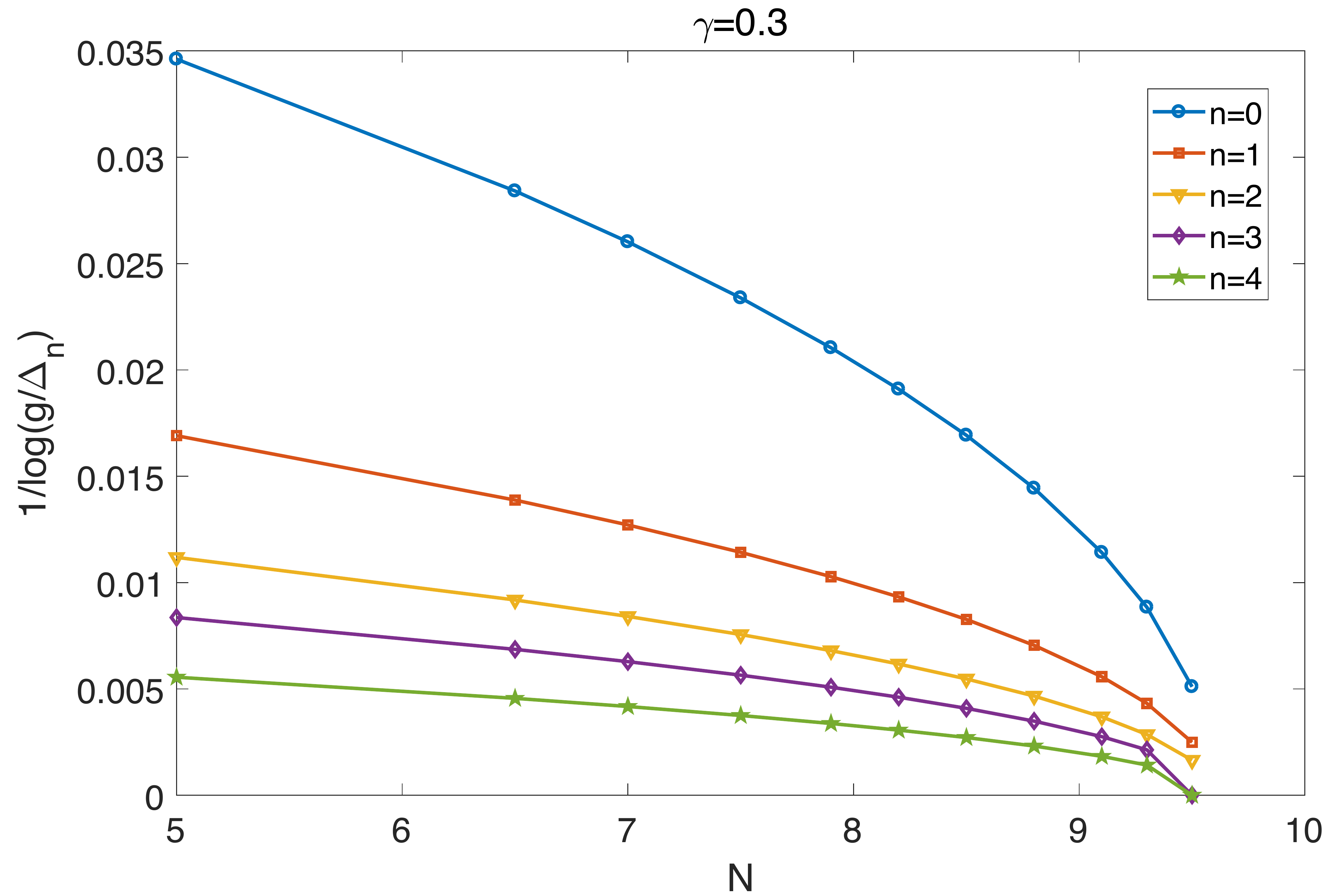}
       \caption{The values of the gap function at zero frequency, $\Delta_n$ for solutions with different $n=0,1,..4$ for representative $\gamma=0.3$. Observe that all $\Delta_n$ vanish at $N = N_{cr}$.}
       \label{fig:Phi0n}
     \end{center}
  \end{figure}

  Below we present a self-consistent reasoning how one can find a solution of the non-linear gap equation.  Namely, we assume that $\Phi (x)$ can be approximated by a constant $\Phi_0$ up to some $x = x^*$, and  at larger $x$ reduces to the solution of the linearized gap equation.  This sets up three conditions: (i)  $\cos(\beta \log{x^*} + \phi) = \sqrt{x^*}$, (ii)  $\Sigma (x^*) = \Phi_0$, i.e., $\Phi_0 = \omega_0 (x^*)^{(1-\gamma)/\gamma}$, and (iii) $x^* =0$ for $\beta =0$.   The first equation determines $x^*$, the second relates the magnitude of $\Phi_0$ to $x^*$, and the third implies that a non-zero $\Phi_0$ is only possible at $N < N_{cr}$, when $\beta >0$.  We remind that the phase $\phi$ is already fixed at some certain value in the interval $[0,\pi/2]$, i.e., $x^*$ is the only unknown. Solving the first equation at small $\beta$ (i.e., at $N \leq N_{cr}$) we find  an infinite discrete set of solutions $x^*_n = Q e^{-n\pi/\beta}$, where $Q \approx e^{(\pi/2- \phi)/\beta}$ and $n=1,2,3...$.  Accordingly, there is discrete set of the gap magnitudes $\Delta_{0,n} = \Delta_n \propto \omega_0 e^{-n \pi \beta (1-\gamma)/\gamma}$.  We show different $\Delta_n$  in Fig.\ref{fig:Phi0n}.

   The implication is that $N = N_{cr}$ is a very special critical point:  on one side of this point, at $N > N_{cr}$, the system is frozen at the onset of gap opening, on the other side, at $N < N_{cr}$, the system develops an infinite set of solutions of the non-linear gap equation. The end point of the set is the solution of the linearized gap equation.  These solutions are topogically distinct in the sense that  $\Phi_n (\omega)$ changes sign $n$ times as a function of frequency before saturating at the value $\Phi_{0,n}$ at the smallest $\omega_m$. The largest gap magnitude is for the solution with $n=0$, for which $\Phi_0 (\omega_m)$ does not change sign.

A numerical verification of the existence of an infinite set of solutions at $T=0$ requires extra efforts,
  because numerical calculations are normally done for a finite number of discrete Matsubara frequencies.
 However, by a simple logics, each solution  $\Phi_n (\omega_m)$  should vanish at its own $T_{p,n}$, whose existence can be verified by solving the linearized gap equation in different topological sectors. This is what we do in the next section.

 Before we go to finite $T$, a few remarks about $T=0$. First,
   our argument that the solutions of the linearized gap equation excists for all $N$
 is appealing, but still approximate because we converted the original  integral equation into a second order differential equation.  As the full proof, in  Ref.\cite{ac_exact} we obtained the exact solution of the linearized gap equation, valid for all $\gamma <1$ and all $N$.  At small $\gamma$ the exact solution is quite similar to the one we presented above, at larger $\gamma$ there are quantitative, but not qualitative differences.

Second, $N_{cr}$ can be obtained for any $\gamma$, not necessary small, by analyzing  the power-law solution of the linearized gap equation at small $\omega$ and checking when the exponents change from real to complex. For arbitrary $\gamma <1$  we obtain
\beq
N_{cr} =\frac{\frac{\pi}{2} (1-\gamma)}{\sin{\frac{\pi}{2} (1-\gamma)}} \frac{\pi}{\Gamma (\gamma)} \frac{\left(1 - \cos{\frac{\pi \gamma}{2}}\right)^{-1}}{\Gamma^2\left(1-\gamma/2\right)}
\label{su_15_1}
\eeq
\begin{figure}
  \begin{center}
    \includegraphics[width=7cm]{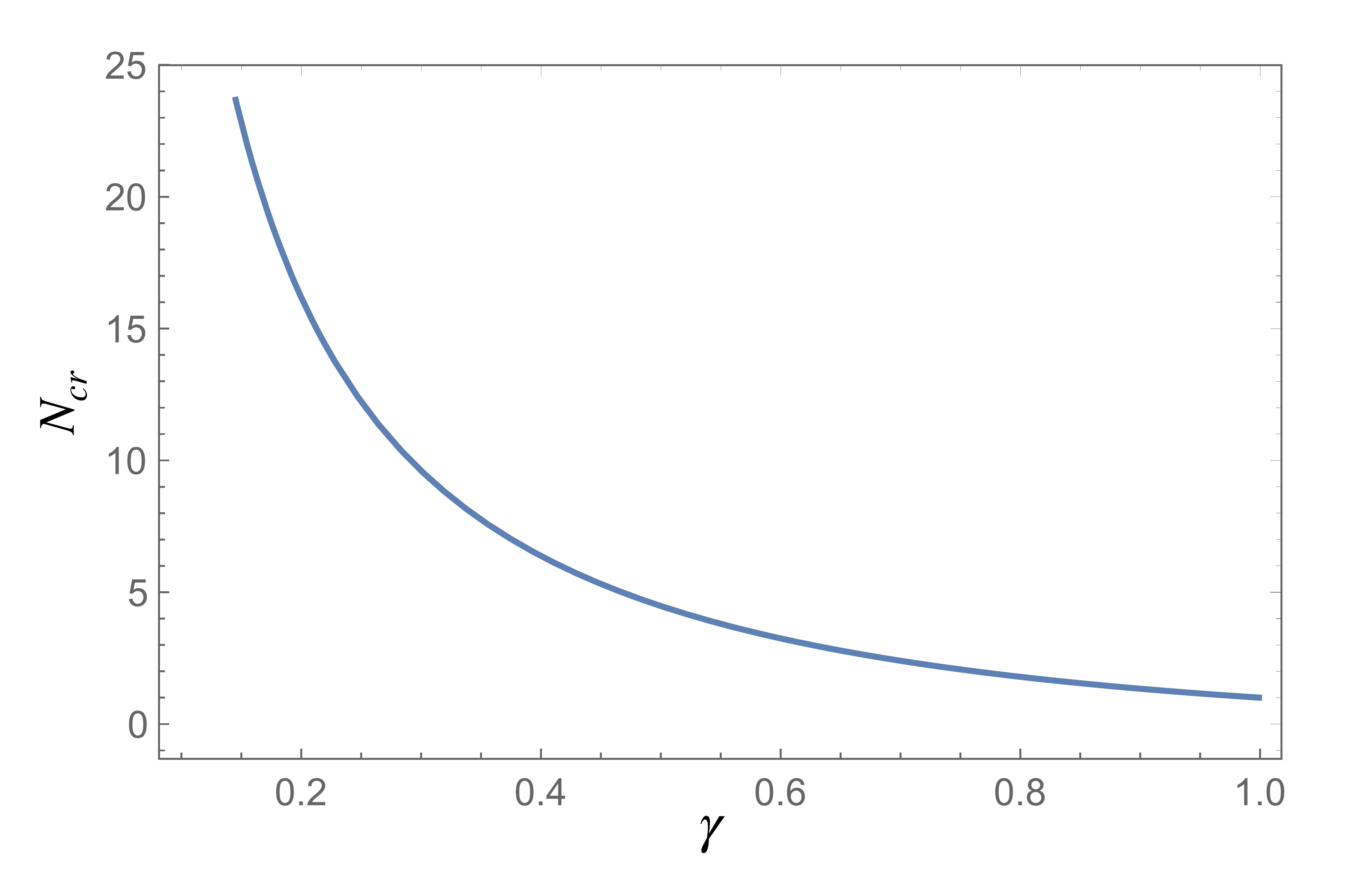}
    \caption{The critical value of $N$ at $T=0$ -- $N_{cr}$, as a function of $\gamma$ (see Eq.
     \eqref{su_15_1}). The system is a critical NFL at $N > N_{cr}$ and a superconductor at $N < N_{cr}$. }\label{fig:Ncr}
  \end{center}
\end{figure}
We plot $N_{cr}$ vs $\gamma$ in Fig.\ref{fig:Ncr}
Eq. (\ref{su_15_1}) has been obtained in Ref. \cite{acf} for $\gamma =1/2$, Ref. \cite{raghu_15} for small $\gamma$, and Ref. \cite{Wang2016} for arbitrary $\gamma$.
  A similar result has been recently found in the study of the pairing in the SYK-type model ~\cite{wang_19,schmalian_19,schmalian_19a}  (see the article by Daniel Hauck, Markus Klug, Ilya Esterlis, and J\"{o}rg Schmalian for this issue).

\section{Multiple solutions for the onset temperature of the pairing}
\label{finite_T_2}
\begin{figure*}
  \begin{center}
    \includegraphics[width=16cm]{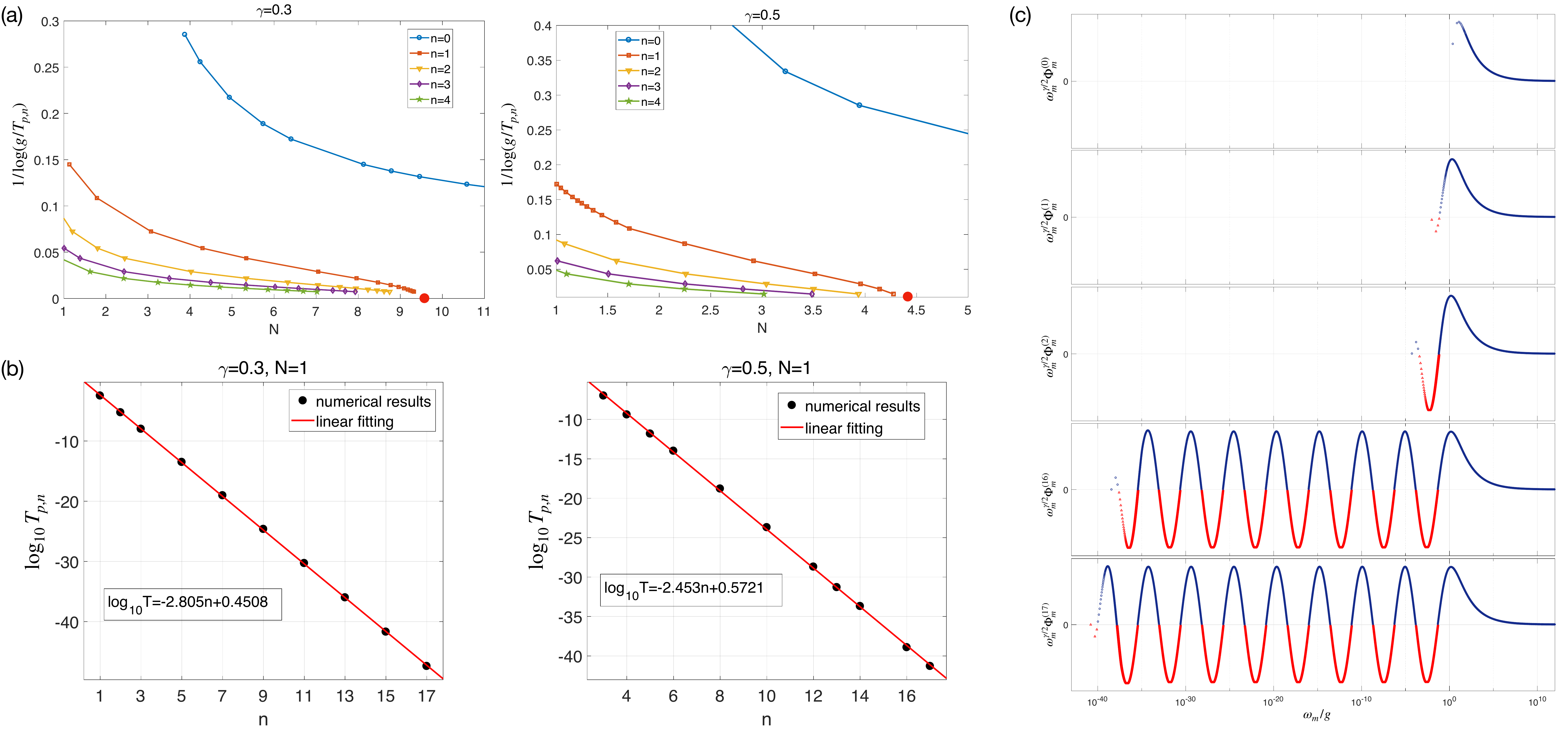}
    \caption{Multiple solutions for the onset temperature of the pairing, $T_{p,n}$ ($n=0,1,2,3..$)  (a) $T_{p,n}$ for $n =0, 1..4$  for $\gamma=0.3$ and $\gamma =0.5$. Observe that all $N_{p,n}$ vanish at $N = N_{cr}$ (red dots in (a)).  The largest $T_{p,0}=T_p$ bypasses $N_{cr}$; (b) Plots of $\log(T_{p,n})$ vs. the index $n$. The linear dependence with a negative slope indicates that $T_{p,n}$ decay exponentially with $n$. For definiteness we set $N=1$; (c)Plots of $\Phi_n (\omega_m)$ at $T = T_{p,n}$ for solutions with $n =0,1,2,  16$, $17$.  We set $\gamma=0.5$ and $N=1$. We see that $\Phi (\omega_m)$ oscillates on a logarithmic scale, and the $n$th solution changes sign $n$ times as function of Matsubara frequency $\omega_m$. The solution with $n \to \infty$, for which $T_{p,n} \to 0$, changes sign infinite number of times, like the $T=0$ solutions in Fig. \ref{fig:Phix}.}\label{fig:Tpn}
  \end{center}
\end{figure*}
In Sec.  \ref{Tp} we found numerically  the onset temperature of the pairing $T_p$. We now show that
this is the largest temperature of the set $T_{p,n}$ of onset temperatures for topologically different solutions. In this set $T_{p,0} = T_p$ and $T_{p,n}$ at $n \to \infty$ tends to zero.  To shorten the presentation we show the numerical evidence (Ref. \cite{WAWC}).

In Fig.\ref{fig:Tpn}(a) we show $T_{p,n}$ obtained by analyzing the eigenvalues of the linearized equation for $\Phi (\omega_m)$ for a certain $\gamma$.  We clearly see that there is infinite set of non-zero $T_{p,n}$.
The largest $T_{p,0} = T_p$ is different from all other $T_{p,n}$ in that it remains finite for all $N$.
 All other $T_{p,n}$ vanish at $N = N_{cr}$, as evidenced from the Fig.\ref{fig:Tpn}(a)
 Because both $\Phi_n (\omega)$ and $T_{p,n}$ for $n \geq 1$  vanish at $N= N_{cr}$, it is natural to expect that they are of the same order.  By this argument, $T_{p,n} \propto \omega_0 e^{-\beta n (1-\gamma)/\gamma}$, i.e., $\log{T_{p,n}/\omega_0}$ scales linearly with the number of the solution, $n$.
  In Fig.\ref{fig:Tpn}(b) we plot $T_{p,n}$ in a logarithmic scale. We clearly see that $\log{T_{p,n}/\omega_0}$ is a linear function of $n$, as we anticipated.

   In Fig.\ref{fig:Tpn}(c) we show $\Phi_n (\omega_m)$ for different solutions. Each solution is plotted vs a discrete frequency $\omega_m = \pi T_{p,n}  (2m +1)$. The smallest $\omega_0 =\pi T_{p,n}$ gets progressively smaller with increasing $n$. We see that, as we expected, $\Phi_n (\omega_m)$ changes sign $n$ times.
    This is fully consistent with the solution for a finite $\Phi_n (\omega)$ at $T=0$.

\begin{figure}
  \begin{center}
    \includegraphics[width=7cm]{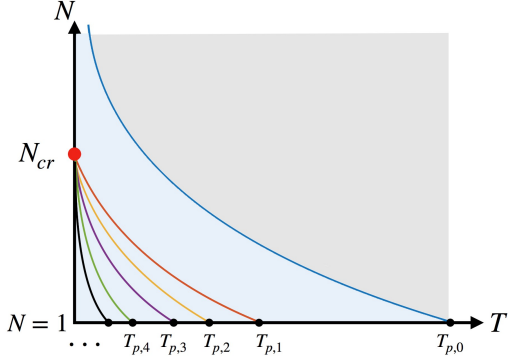}
    \caption{Cartoon of the  behavior of $T_{p,n}(N)$ vs $N$ for arbitrary  $\gamma<1$. All $T_{p,n}$ with $n >0$ terminate at $N = N_{cr}$, while $T_{p,0} = T_p$ remains non-zero for any $N$. }\label{fig:TpnN}
  \end{center}
\end{figure}

      The outcome of this analysis is shown in Fig.\ref{fig:TpnN} -  there exists an infinite set of the lines $T_{p,n} (N)$ for $n >0$, which all terminate at $T=0, N=N_{cr}$, and a standalone line $T_{p,0} (N) = T_p (N)$, which does not terminate at any $N$.

\section{The origin of $T_{\crr} (N)$}
\label{final}

 We now relate the existence of multiple lines $T_{p,n} (N)$ representing different solutions of the linearized Eliashberg equation and the crossover line
  $T = T_{\crr}$, which we observed in Sec. \ref{sec:smallerN}  by solving numerically the non-linear gap equation for larger and smaller $N$.  First, we naturally identify the end point of $T_\crr (N)$ in Fig.\ref{fig:phasediagram}
  with $N_{cr}$, which we found in the $T=0$ analysis. Second, the largest
   condensation energy at $T=0$ corresponds to the solution with $n=0$. This solution is the only global minimum of the Free energy. Other solutions are local minima.
   This also holds at a finite $T$.   In this respect, $T_{p,0} = T_0$ is the only
  onset temperature for the pairing.
  However, the  functional form of the gap function $\Delta_0 (\omega_m, T)$ evolves with decreasing $T$ because  other gap components also get generated below $T_p$ because of non-linear coupling in the Free energy between $\Delta_0 (\omega_m, T_p)$ and  $\Delta_n (\omega_m, T_{p,n})$ with $n >0$ (Refs.\cite{Yang2000,avi_1})
  As a result, as $T$ decreases, not only the magnitude of the actual gap function $\Delta_0 (\omega_m)$ get larger, but its frequency dependence also changes.
   Near $T_p$, the relative weight of $n >0$ components is small  and $\Delta_0 (\omega_m, T) \propto (T_p-T)^{1/2}  \Delta_0 (\omega_m, T_p)$.    This is the regime of ``gap filling'' behavior.
   At smaller $T$, the weight of $n >0$ components in $\Delta_0 (\omega_m, T)$  increases, and eventually
   $\Delta_0 (\omega_m, T)$  becomes a smooth function of $\omega_m$,  as, we found, it is at $T=0$. As we found in Sec. \ref{sec:smallerN}, for such form of $\Delta_0 (\omega_{m}, T)$ the system displays BCS-type ``gap closing'' behavior.

These reasoning show that the existence of multiple $T_{p,0} (N)$  for the solution of the linearized gap equation is crucial for the existence of the crossover from non-BCS ``gap filling'' to BCS-like ``gap closing''  behavior. The functional form
 of $T_{\crr} (N)$ is a more subtle issue, which we do not address here.  Our reasoning is valid for $N \leq N_{cr}$, where $T_{\crr}$ and $T_{p,n}$ for $n >0$ are all small (they all vanish at $N = N_{cr}$).  Numerical results show that at smaller $N$,
  $T_{\crr} (N)$ becomes numerically larger than the largest of  $T_{p,n}$.

\section{Superfluid stiffness}
\label{rho}
 So far we found that at $T_{\crr} < T < T_p$, the feedback from the pairing on fermions is weak, i.e., fermionic self-energy retains its NFL form and
   the system displays "gap filling" behavior. This result does not address whether or not the system has long-range phase coherence.  It is natural to ask how strong  phase fluctuations are in the range $T_{\crr} < T < T_p$.

 Superfluid stiffness has been computed in Ref. \cite{abanov_last} by expressing the coordinate-dependent gap function as $\Delta (\omega_m,r) = \Delta (\omega_m) e^{i \phi (r)}$ and evaluating the term in the effective action $\int dr (\nabla \phi (r))^2$. The stiffness
  $\rho_s (T)$ is the prefactor in this term.   For a BCS superconductor  $\rho_s (T=0) \approx E_F/(4\pi)$. Because $E_F$ is assumed to be much larger than mean-field  transition temperature $T_p$, phase fluctuations are weak and mean-field  $T_p$ almost coincides with the actual $T_c$.
   In our case we found at large $N$, when $T_{\crr} =0$,
      \beq
    \rho_s (T) \approx \frac{T}{N} \left(1- \left(\frac{T}{T_p}\right)^\gamma\right)  \frac{E_F}{\pi T \chi (0)} \left(1 + O\left(\frac{1}{N}\right)\right),
    \eeq
    where $\chi (0)$ is a static susceptibility of a critical bosonic field.  Formally, $\chi (0)$ diverges at a QCP.  However, setting $\chi (0)^{-1}$ to zero would invalidate  Eliashberg  theory, which is
     built on the notion that there is a small parameter, which makes vertex corrections small and simultaneously allows one to factorize momentum integration by separating  fast electrons and slow bosons.  One can verify that in our case this Eliashberg parameter is
    $E_F/\pi T \chi (0)$.  The consideration based on Eliashberg equations is valid when, at most, $\pi T \chi (0) \geq   E_F$.
       This bounds   $\rho_s (T)$ from above by  $(T/N) \left(1- \left(T/T_p\right)^\gamma\right)$.  We see that $\rho_s (T)$ is at most of order $T/N$, i.e., $\rho_s < T$.  In this situation, phase fluctuations are strong, $\langle \phi^2 \rangle \geq 1$,
    and long range phase coherence likely destroyed
    ~\cite{emery_kiv,*mohit_1,benfatto}.   Applying this reasoning to smaller $N$, we find that $\rho_s \leq T$ at $T$ where the pairing is induced by fermions with $\omega_m = \pm \pi T$, i.e., in the range
      $T_{\crr} < T < T_p$.
     Then, at least a portion of this range is actually phase-disordered, i.e., the actual $T_c$ is of order $T_{\crr}$.
       At smaller $0< T < T_{\crr}$  the same calculation yields,
       $\rho_s \geq \Delta (T \to 0) \geq T_{\crr}$, i.e., phase fluctuations are weak and phase coherence survives.  The outcome of this analysis is that the region below $T_{\crr}$ corresponds to a true SC state, while in the range  $T_{\crr} < T < T_p$ the system displays pseudogap behavior, meaning that the amplitudes of the pairing vertex and the gap function are given by Eq. (\ref{s_20})   and the DOS is given by Eq. (\ref{s_21}), but there is no true long-range order.

\begin{figure}
   \begin{center}
     \includegraphics[width=8cm]{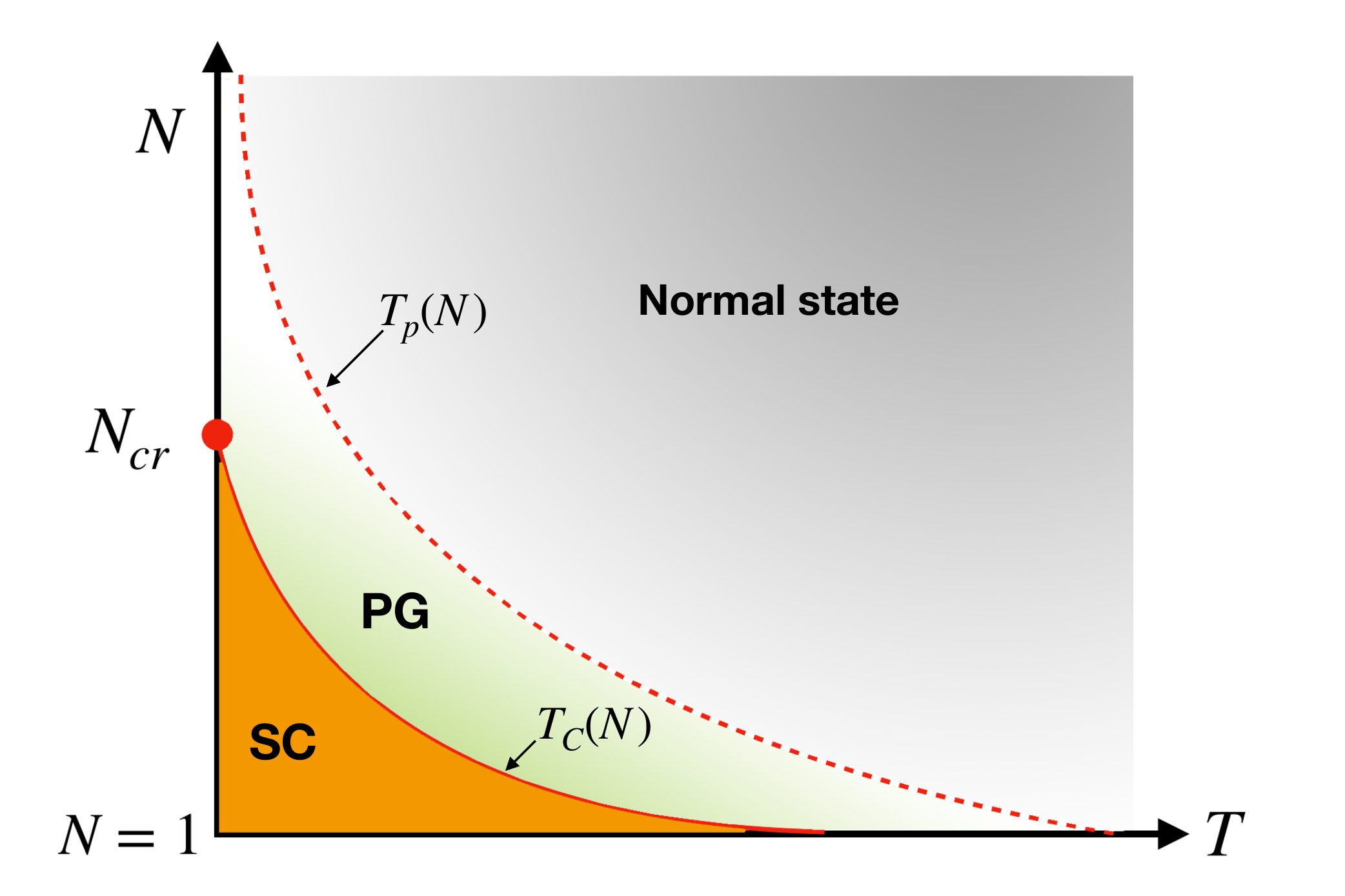}
     \caption{Phase diagram of the $\gamma$ model with $0 < \gamma <1$, emerging from our analysis.
       The red solid line is the actual transition temperature
      $T_c (N)$. Below $T_c (N)$ the system has long-range superconducting order, and the observables display BCS-like behavior.  The dashed red line marks the crossover temperature $T_p(N)$. Below this temperature the system displays "gap filling" behavior, which we described in Sec. \ref{finiteT_1}, but there is no true long-range order. The range between $T_c (N) $ and $T_p (N)$ is a portion of the pseudogap phase, where pairing fluctuations are strong. In the cuprates, pseudogap behavior likely persists above $T_p (N)$ due to other effects, which we didn't consider here. }\label{fig:phasediagram_1}
   \end{center}
 \end{figure}

       We show the resulting phase diagram in Fig. \ref{fig:phasediagram_1} It is similar to Fig. \ref{fig:phasediagram}, but the former crossover line $T_{\crr}$ is the actual $T_c$ line (the solid line in the figure), while $T_p$ is the crossover temperature, below which the behavior of the gap function and the DOS is the
        same as of our $\Delta (\omega_m)$ and $N(\omega_m)$, but there is no true superconducting order.
        The region between $T_p$ and $T_c$  is called pseudogap phase, or, in our case, a precursor to superconductivity.

\section{Application to the $d$-wave case}
\label{sec:application}
\begin{figure*}
  \begin{center}
    \includegraphics[width=14cm]{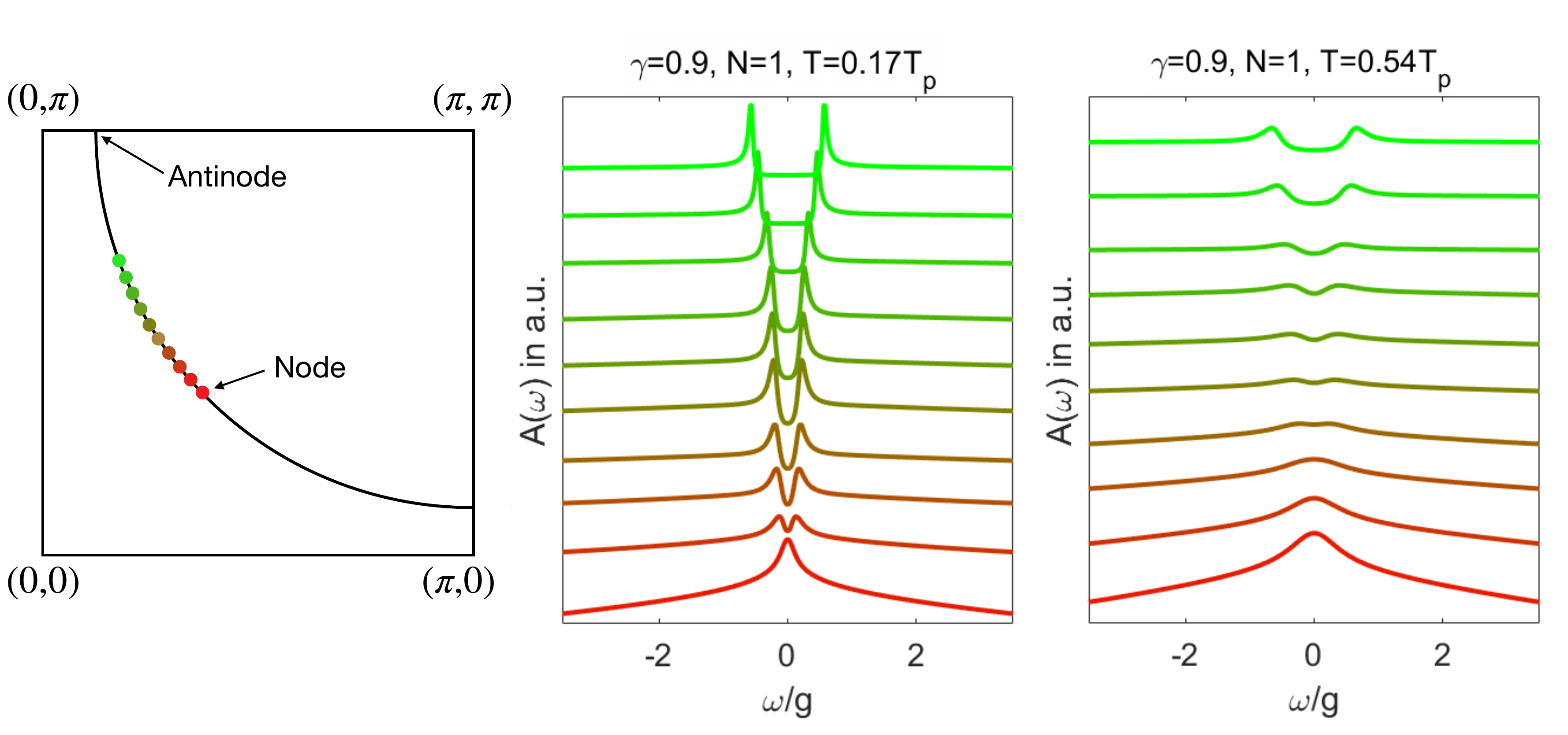}
    \caption{The spectral function for a $d-$wave superconductor, $A_k(\omega)$, along the Fermi surface at  $T<T_{\crr}$ and $T>T_{\crr}$. The nodal and anti-noidal regions are denoted by red and green colors, respectively. At low $T$, $A_k(\omega)$  has two peaks,  which merge at the nodal point. At higher $T$, $A_k(\omega)$ in the nodal region develops a single peak at $\omega =0$. The region where this happens is called a Fermi arc. In the antinodal region the peaks persist get "filled in" when the temperature increases towards $T_p$. }\label{fig:cuprates}
  \end{center}
\end{figure*}
Finally, we briefly discuss the relation between our results and ARPES data for cuprate superconductors.
To quantitatively apply our results to the cuprates, we
 (i) assumed that the critical boson is a $(\pi,\pi)$ spin fluctuation, (ii)
  modeled the $d-$wave symmetry of the gap function
     by adding
     $\cos {2\theta}$ factor to $\Phi^* (\omega)$, and (iii) used as an input the fact that spin fluctuations become nearly propagating modes below $T_p$ due to the feedback from
       the pair formation on bosonic self-energy~\cite{finger_2001}, in which case the exponent $\gamma \leq 1$.
        In Fig.\ref{fig:cuprates}
         we show the results for the spectral function  $A_{{\bf k}_F} (\omega)$
       for ${\bf k}_F$ in near-nodal and anti-nodal regions.
        The difference between the two is partly due to $d-$wave gap symmetry and partly due to
        the  difference in the contribution from thermal fluctuations,
       which are much stronger in the antinodal region.
      We see that at $T < T_{\crr}$,  $A_{{\bf k}_F}(\omega)$ has two
         peaks, more strongly separated in the antinodal region. This is an expected result for a $d$-wave BCS-like superconductor.
      At higher $T > T_{\crr}$, $A_{{\bf k}_F}(\omega)$ near the nodes has a single maximum at $\omega =0$, while in the antinodal region $A_{{\bf k}_F}(\omega)$ has a dip at $\omega =0$ and a shallow maximum, whose frequency scales with $T$ (the "gap filling").   This behavior  reproduces the key features of ARPES data detected in Refs.\cite{dessau,*dessau_1,*dessau_2,*dessau_3,kanigel,kaminski,*Kaminski2,shen,*shen2,shen3,shen4,*kordyuk2,hoffman,Peng2013}.
        The behavior of $N(\omega)$ is quite similar to that of $A(\omega)$ in the antinodal region.   This is fully consistent with the  STM data~\cite{DOS,hoffman}.

   We emphasize that in our analysis we only considered fluctuations in the particle-particle channel and ignored another aspects of pseudogap phase, such as precursor to Mott/antiferromagnetic phase, or a development of a competing order in the particle-hole channel.   In this respect, our reasoning is applicable only to a portion of a pseudogap phase, where pairing correlations are strong~\cite{ong2010}.

\section{Summary}
\label{sec:summary}
In this mini-review, we used Eliashberg theory to analyze the interplay between NFL and SC near a quantum-critical point in a metal.
 We considered  a class of quantum-critical models with an effective dynamical electron-electron interaction
 $V(\Omega_m) \propto 1/|\Omega_m|^\gamma$ (the $\gamma$-model) for $0 < \gamma <1$.
  We argue that the tendency towards pairing is stronger, and the ground state is a superconductor. We argue, however, that there exist two distinct regimes of system behavior below the onset temperature of the pairing $T_p$. In the range $T_{\crr} < T < T_p$ fermions remain incoherent, and the spectral function $A(k, \omega)$ and the DOS  $N(\omega)$  both display "gap filling" behavior, meaning that, e.g., the position of the maximum in $N(\omega)$ is set by temperature rather than the pairing gap. At lower $T < T_{\crr}$, fermions acquire coherence, and $A(k, \omega)$ and  $N(\omega)$  display BCS-like "gap closing" behavior.  We argue that the existence of the two regimes comes about because of special behavior
   of fermions with frequencies $\omega = \pm \pi T$  along the Matsubara axis. Specifically, for these fermions, the component of the self-energy, which competes with the pairing, vanishes in the normal state.  We further argue that the crossover  at $T \sim T_{\crr}$ comes about because Eliashberg equations allow an infinite number of topologically distinct solutions for the onset temperature of the pairing within the same gap symmetry. Only one solution, with the highest $T_p$, actually emerges, but other solutions are generated and  modify the functional form of the gap function
    at around $T_{\crr}$. Finally, we argue that the actual $T_c$ is comparable to $T_{\crr}$, while at $T_{\crr} < T < T_{p}$ phase fluctuations destroy superconducting long-range order.

 \acknowledgements
  We thank  B. Altshuler, E. Andrei,  E. Berg, A. Bernevig, L. Classen, P. Coleman, R. Combescot, D. Dessau, D. Haldane, I. Esterlis, R. Fernandes, A. Finkelstein,  B. Keimer, A. Kanigel, S. Kivelson, I. Klebanov, A. Klein, G. Kotliar, S. Lederer, D. Maslov, A. Millis, V. Pokrovsky, N. Prokofiev, P. Ong, S. Raghu, M. Randeria, S. Sachdev, Y. Schattner, S. Sondi, J. Schmalian, G. Torroba, A-M Tremblay, A. Yazdani, E. Yuzbashyan, and J. Zaanen for useful discussions.   The work by  AVC and YW was supported by the NSF  DMR-1834856.

\bibliography{eliashberg90}
\end{document}